\documentclass[fleqn,usenatbib]{mnras}

\usepackage{newtxtext,newtxmath}

\usepackage[T1]{fontenc}

\usepackage{graphicx}   
\usepackage{amsmath}    
\usepackage{pdflscape}
\usepackage{longtable}

\title[Optical Study of TRAPUM Pulsars]{Optical Study of TRAPUM Pulsars and Modelling of the Redbacks: PSR~J1036$-$4353 and PSR~J1803$-$6707}

\author[A. Phosrisom et al.]{
A. Phosrisom,$^{1,2}$\thanks{E-mail: adipol.phosrisom@manchester.ac.uk; adipol@narit.or.th}
R. P. Breton,$^{1}$
C. J. Clark,$^{3,4}$
M. Burgay,$^{5}$
J. Strader,$^{6}$
L. Chomiuk,$^{6}$
K. V. Sokolovsky,$^{7}$\newauthor 
I. Molina,$^{6}$
R. Urquhart,$^{6}$
M. R. Kennedy,$^{8,1}$,
S. J. Wagner,$^{9}$
V. S. Dhillon,$^{10}$
O. G. Dodge,$^{1,3}$\newauthor
B. W. Stappers,$^{1}$
T. Thongmeearkom$^{1,2}$
\\
$^{1}$ Jodrell Bank Centre for Astrophysics, The University of Manchester, Manchester, UK\\
$^{2}$ National Astronomical Research Institute of Thailand, Don Kaeo, Mae Rim, Chiang Mai 50180, Thailand\\
$^{3}$ Max Planck Institute for Gravitational Physics (Albert Einstein Institute), D-30167 Hannover, Germany \\
$^{4}$ Leibniz University Hannover, D-30167 Hannover, Germany \\
$^{5}$ INAF - Osservatorio Astronomico di Cagliari, via della Scienza 5, 09047 Selargius (CA), Italy\\
$^{6}$ Center for Data Intensive and Time Domain Astronomy, Department of Physics and Astronomy, Michigan State University, East Lansing, MI 48824, USA\\
$^{7}$ Department of Astronomy, University of Illinois at Urbana-Champaign, 1002 W. Green Street, Urbana, IL 61801, USA\\
$^{8}$ School of Physics, Kane Building, University College Cork, Cork, Ireland \\
$^{9}$ LSW, ZAH, Heidelberg University, Koenigstuhl 12, 69117 Heidelberg, Germany\\
$^{10}$ Department of Physics and Astronomy, University of Sheffield, Sheffield S3 7RH, UK\\
}

\date{Accepted 2025 December 4. Received 2025 November 27; in original form 2025 March 14}

\pubyear{2025}

\begin{document}
\label{firstpage}
\pagerange{\pageref{firstpage}--\pageref{lastpage}}
\maketitle

\begin{abstract}
The Transients and Pulsars with MeerKAT (TRAPUM) project discovered eight binary millisecond pulsars in its first shallow \textit{L}-band survey of unidentified \textit{Fermi} $\gamma$-ray sources using the MeerKAT radio telescope. We conducted follow-up observations using ULTRACAM on the New Technology Telescope at the La Silla Observatory to search for the optical counterpart to the pulsar companions. We found two redback companions, in PSRs J1803$-$6707 and J1036$-$4353, and provided upper limits for the other pulsar binaries. We used the \texttt{Icarus} code to fit the redback's light curves using various irradiation models. The asymmetric double-peak light curves of PSR~J1036$-$4353 are best fit with diffusion and convection models. Comparing the two prescriptions of irradiation and gravity darkening, models with post-irradiation gravity darkening provide superior fits (particularly for lower gravity-darkening exponents), suggesting that the irradiation energy is deposited deep in the stellar photosphere. PSR~J1803$-$6707, on the other hand, displayed variability in the amplitude of its irradiation-dominated light curves over a time scale of a few months. This effect can be modelled only if the companion's filling, irradiation temperature, and convection coefficients are allowed to vary over time. Had the star been closer to filling its Roche lobe, like in the cases of the known transitional millisecond pulsars J1023+0038 and J1227$-$4853, this 4.1 per~cent variation in the volume-averaged filling of the star would have caused it to experience a state change to form an active accretion disc.
\end{abstract}




\section{Introduction}
The population of known millisecond pulsars (MSPs) with rotational periods ($P_{\rm spin} <$ 15 ms) exceeds 500, constituting over 15 per cent of the total number of identified pulsars, which currently stands at more than 3800 (ATNF catalogue v2.6.0; \citealt{Manchester2005}). Over 300 MSPs exist in binary systems, typically paired with degenerate stars such as white dwarfs or low-mass stars. Most MSP binaries having low-mass companion stars in tight orbit ($P_{\rm orb}$ $\la$ 1 day), often referred to as `spider' pulsars, are strongly influenced by the intense pulsar irradiation. These pulsars are further categorised according to the companion mass into two sub-classes: \textit{black widows} ($M_c \sim 0.01-0.07 M_{\sun}$) and \textit{redbacks} ($M_c \sim 0.3-0.7 M_{\sun}$; \citealt{DAmico2001,Roberts2012,Swihart2022}). It is generally thought that these binaries are the result of a recycling process during which the pulsars are spun up through mass accretion while their high-energy radiations evaporate the companions \citep{Alpar1982, Fruchter1990, Bhattacharya1991}. The evolutionary paths and connections between these sub-classes remain subjects of ongoing debate. \citet{Chen2013} proposed that black widows and redbacks are two different groups with their evolutionary tracks determined by evaporation efficiency while \citet{Benvenuto2014} suggested that some redbacks with short orbital period ($P_{\rm orb} \la$ 6 hours) will evolve to black widows while the others will become pulsar-helium white dwarf binaries. \citet{Misra2025} also confirmed that some redbacks will evolve into black widows.

Transitional millisecond pulsars (tMSPs) exhibit back-and-forth transitions between a state characterised by mass transfer and the presence of a disc, and another one in which there is no disc and no mass transfer on time-scales of years. They serve as a direct connection between low-mass X-ray binaries (LMXBs) and MSP binaries, thereby substantiating the recycling scenario \citep{Papitto2022}. As of now, such transitions have been identified in three systems, namely PSR~J1023+0038 \citep{Archibald2009, Patruno2013, Stappers2014}, PSR~J1824-2452I/IGR J18245-2452 \citep{Papitto2013}, and XSS J12270-4859 \citep{Roy2015}. These systems are very similar to redback MSPs while in their disc-less `pulsar' state. Proposed mechanisms governing these transitions include irradiation feedback \citep{Benvenuto2014} and thermal-viscous instability in accretion disc \citep{Jia2015}. In both cases, Roche-lobe overflow (RLOF) is expected to cause the formation of the accretion disc.

Optical light curves of spider pulsars feature two main sinusoidal modulation components, one at the orbital frequency due to pulsar-facing irradiation, and the other at twice the orbital frequency due to ellipsoidal modulation from tidal distortion \citep{Breton2011, Schroeder2014}. The relative importance of these effects vary, especially in redbacks \citep[see review in ][]{Turchetta2023}. Strong irradiation results in single-humped light curves. In contrast, on less irradiated redbacks, ellipsoidal modulation due to tidal distortion becomes more significant, creating double-peaked light curves. Beyond the interplay of irradiation and the ellipsoidal modulation, asymmetrical features manifest in the light curves of numerous spiders, including PSR~J1023+0038 \citep{Stringer2021}, PSR~J1048+2339 \citep{Cho2018, Yap2019, Zanon2021}, PSR~J1628$-$3205 \citep{Li2014}, PSR~J1810+1744 \citep{Schroeder2014, Romani2021}, PSR~J2039$-$5617 \citep{Salvetti2015}, and PSR~J2215+5135 \citep{Linares2019, Romani2016, Voisin2020}. Various models, including intra-binary shock \citep{Romani2016}, magnetic ducting spots \citep{Sanchez2017}, and heat redistribution on stellar atmospheres \citep{Kandel_2020_atmo_circ, Voisin2020, Stringer2021}, were proposed to explain these asymmetries.

Over a time span exceeding a decade, the $\gamma$-ray full sky survey performed with the Large Area Telescope \citep[LAT;][]{Atwood2009} aboard the \textit{Fermi} $\gamma$-ray Space Telescope has unveiled positions of potential new energetic pulsars, a significant portion of which are MSPs and spider pulsars \citep{Smith2023}. This offers an effective approach for radio surveys aiming at these positions. The TRAPUM collaboration \citep{Stappers2018} utilised this strategy to search for new pulsars between 2020 and 2021 \citep{Clark2023Lband}. In the course of this survey, we successfully identified nine new MSPs, including two redback binaries. This paper focuses on the concurrent optical follow-up observations we conducted using ULTRACAM aimed at identifying the optical counterparts of the newly discovered pulsar companions and acquiring comprehensive phase-resolved photometric data for the redback pulsars. In addition to the optical follow-up observations, the collaboration also ran radio and $\gamma$-ray timing campaigns for these sources, which could be aided by the precise localisation of the optical counterparts and, in some cases, an approximate orbital period when variability can be observed. The detailed findings of these timing studies are presented in \citet{Burgay2024+timing}.

The paper is organised as follows: Section~\ref{sec:observations} describes the observing setup and data processing performed on the optical data; Section~\ref{sec:modelling} describes the light curve modelling of the new redbacks; Section~\ref{sec:discussion} discusses the implications of our modelling on the physics of spiders and their evolution; and Section~\ref{sec:conclusion} presents concluding remarks.

\begin{table*}
    \centering
    \renewcommand{\arraystretch}{1.1}
    \caption{Expected companions, precise positions, orbital period, time of the ascending node, minimum companion mass and distances from dispersion measurements of the MSP binaries from \citet{Burgay2024+timing}.}
    \label{tab:timing_params}
    \begin{tabular}{@{}c|@{}c|@{}c|@{}c|@{}c|@{}c|@{}c|@{}c|@{}c|@{}c}
        \hline
        Source & Expected & RA & Dec & $P_{\rm orb}$ & $T_{\rm asc}$ & Minimum$^{\rm a}$ & $d^{\rm b}$\\
         & Companion & & & (d) & (MJD) & $M_{\rm c}\ (M_{\odot})$ & (kpc) \\
        \hline
        PSR~J1036$-$4353 & Redback & 10:36:30.21513(3) & -43:53:08.7252(5) & 0.25962126(1) & 59536.3052068(15) & 0.23 & 0.4 \\
        PSR~J1526$-$2744 & White Dwarf & 15:26:45.094(7) & -27:44:06.0(3) & 0.2028108283(10) & 59303.205977(5) & 0.08 & 1.3 \\
        PSR~J1623$-$6936 & White Dwarf & 16:23:51.3869(13) & -69:36:49.252(6) & 11.01366617(5) & 59192.916224(4) & 0.19 & 1.3 \\
        PSR~J1757$-$6032 & White Dwarf & 17:57:45.4472(6) & -60:32:12.298(6) & 6.280139736(17) & 59183.4010970(15) & 0.42 & 3.5 \\
        PSR~J1803$-$6707 & Redback & 18:03:04.235339(20) & -67:07:36.1577(4) & 0.380473191(6) & 59409.8403825(6) & 0.29 & 1.4 \\
        PSR~J1823$-$3543 & White Dwarf & 18:23:42.9989(7) & -35:43:40.88(3) & 144.568639(5) & 59091.325976(15) & 0.26 & 3.7 \\
        PSR~J1858$-$5422 & White Dwarf & 18:58:07.7664(6) & -54:22:15.527(7)) & 2.581951179(15) & 59564.9829158(15) & 0.12 & 1.2 \\
        PSR~J1906$-$1754 & White Dwarf & 19:06:14.7894(4) & -17:54:34.31(4) & 6.4910850(3) & 59700.935268(5) & 0.05 & 6.8 \\
        \hline
        \multicolumn{8}{l}{$^{\rm a}$ assuming the pulsar mass of 1.4 $M_{\odot}$} \\
        \multicolumn{8}{l}{$^{\rm b}$ dispersion-measurement distance using the YMW16 electron density model \citep{YMW2016}} \\
    \end{tabular}
\end{table*}

\section{Observations}
\label{sec:observations}

\begin{figure*}
    \centering
    \includegraphics[width=\textwidth]{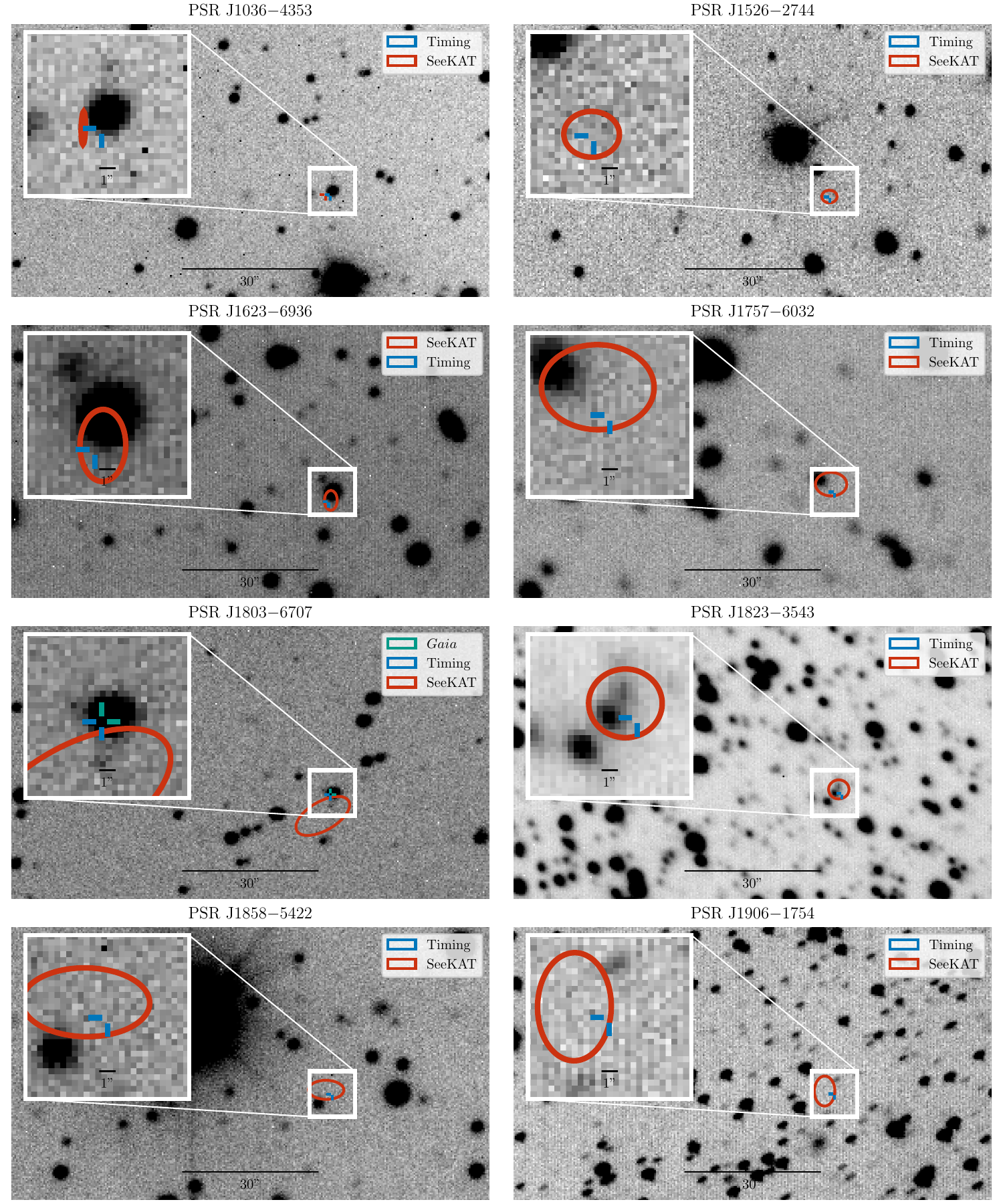}
    \caption{Five-minute stacked ULTRACAM images of the MSP binaries are shown with different localisation ellipses representing 95-per~cent confidence. If the ellipses are smaller than a pixel, they are indicated by two strips pointing at the centres of the ellipses. The SeeKAT localisations are shown in red ellipses. The blue strips indicate the timing localisation obtained during the follow-up timing campaign from \citet{Burgay2024+timing}.}
    \label{fig:loc-plot}
\end{figure*}

\begin{figure}
    \centering
    \includegraphics[width=\columnwidth]{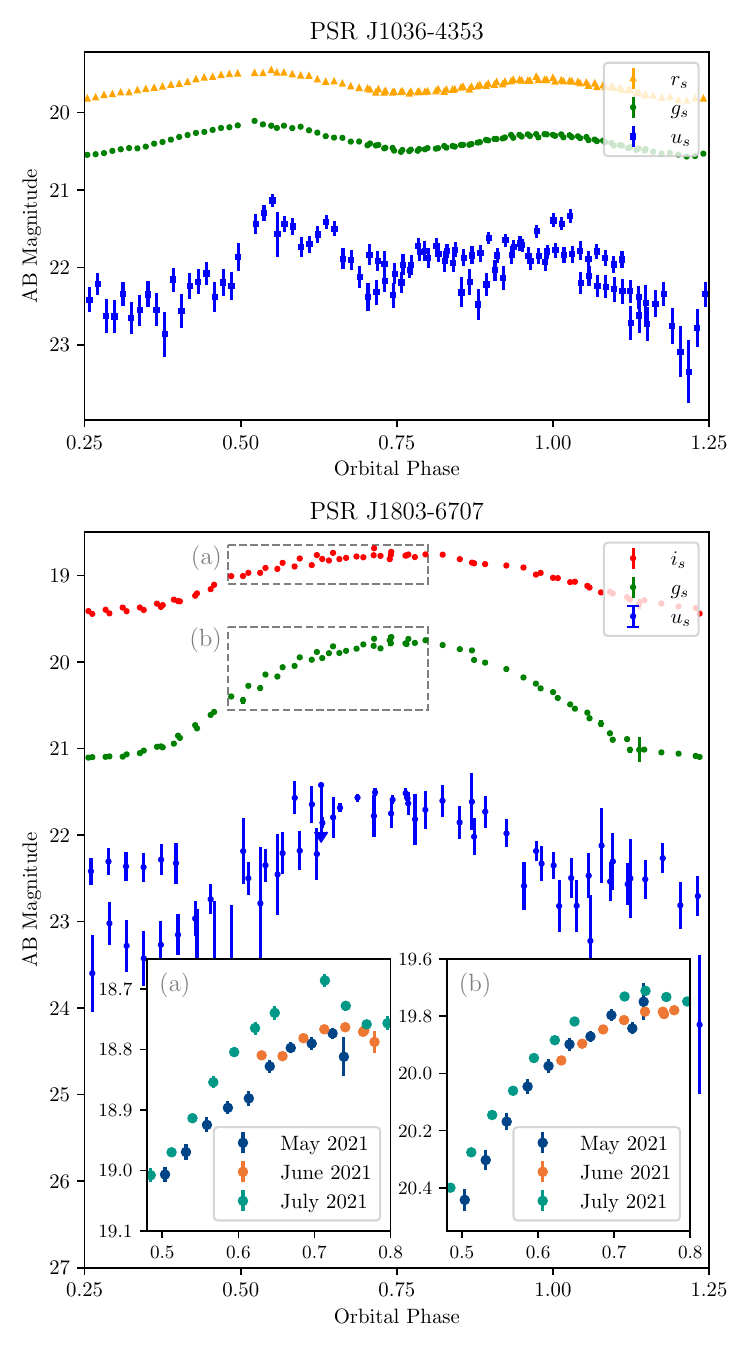}
    \caption{The top panel shows the optical light curves of PSR~J1036$-$4353, observed in $r_s$, $g_s$, and $u_s$ bands. The bottom panel shows the light curves of PSR~J1803$-$6707 in $i_s$, $g_s$, and $u_s$. The data points are re-binned to 300s and 900s for PSRs J1036$-$4353 and J1803$-$6707, respectively. This is to clearly display the discrepancy between observational epochs as shown in the two subpanels (a) and (b). In both panels, if data points fall below the two-sigma detection threshold, the upper limits are shown with downward arrows.}
    \label{fig:first_lc_plot}
\end{figure}

\begin{table*}
    \renewcommand{\arraystretch}{1.1}
    \label{tab:observation_and_upper_limits}
    \caption{Summary of ULTRACAM observations of TRAPUM-discovered sources. The table contains essential details of the observations, including the filters, initiation time of observation, cumulative exposure duration, airmass, average seeing, as well as mean flux and magnitude. If the optical counterpart is not found, the detection limit in radio localisation ellipses is presented instead.}
    \begin{tabular}{@{\:}l|@{\:}c|@{\:}c|@{\:}c|@{\:}c|@{\:}c|@{\:}c|@{\:}c|@{\:}c|@{\:}c|@{\:}c|@{\:}c}
        \hline
        Source & Run Name & Filters & Start MJD & Exposure & Airmass & Airmass & Average & $i_s$ Flux & $i_s$ Mag & $r_s$ Flux & $r_s$ Mag \\
         &  &  &  & (minute) & Start & End & Seeing (\arcsec) & $\mu$Jy &  & $\mu$Jy & \\
        \hline
        PSR~J1036$-$4353 & run011 & $r_s$,$g_s$,$u_s$ & 59671.99428 & 28.6 & 1.235 & 1.166 & 1.40 & - & - & 50(2) & 19.66(3) \\
         & run012 & $r_s$,$g_s$,$u_s$ & 59672.01448 & 285.7 & 1.165 & 1.195 & 1.28 & - & - & 49(4) & 19.7(1) \\
         & run004 & $r_s$,$g_s$,$u_s$ & 59672.99034 & 23.8 & 1.240 & 1.180 & 1.41 & - & - & 58(2) & 19.49(3) \\
         & run005 & $r_s$,$g_s$,$u_s$ & 59673.00718 & 89.2 & 1.179 & 1.054 & 1.40 & - & - & 50(4) & 19.65(9) \\
         & run006 & $r_s$,$g_s$,$u_s$ & 59673.07126 & 129.0 & 1.052 & 1.076 & 1.24 & - & - & 50(3) & 19.66(6) \\
        PSR~J1526$-$2744 & run013 & $i_s$,$g_s$,$u_s$ & 59369.96953 & 16.7 & 1.654 & 1.532 & 1.59 & -5(4) & >21.28 & - & - \\
         & run020 & $i_s$,$g_s$,$u_s$ & 59370.23009 & 11.4 & 1.144 & 1.173 & 1.07 & -1.1(8) & >22.92 & - & - \\
         & run006 & $i_s$,$g_s$,$u_s$ & 59370.95591 & 217.5 & 1.794 & 1.013 & 1.17 & -1(1) & >22.51 & - & - \\
         & run012 & $i_s$,$g_s$,$u_s$ & 59372.96094 & 45.7 & 1.659 & 1.371 & 1.81 & -3(2) & >21.84 & - & - \\
        PSR~J1623$-$6936 & run022 & $i_s$,$g_s$,$u_s$ & 59406.11385 & 37.4 & 1.326 & 1.351 & 1.12 & 5(2) & >21.32 & - & - \\
         & run009 & $i_s$,$g_s$,$u_s$ & 59408.96300 & 17.9 & 1.449 & 1.420 & 1.67 & -1(1) & >22.34 & - & - \\
         & run025 & $i_s$,$g_s$,$u_s$ & 59409.26091 & 15.3 & 1.745 & 1.804 & 1.87 & -3(1) & >22.62 & - & - \\
        PSR~J1757$-$6032 & run009 & $i_s$,$g_s$,$u_s$ & 59407.30654 & 21.1 & 1.546 & 1.634 & 2.15 & 20(10) & >19.73 & - & - \\
         & run016 & $i_s$,$g_s$,$u_s$ & 59409.01972 & 28.0 & 1.349 & 1.291 & 1.68 & 0(1) & >22.54 & - & - \\
         & run029 & $i_s$,$g_s$,$u_s$ & 59409.28807 & 15.1 & 1.479 & 1.533 & 1.50 & 1(1) & >22.34 & - & - \\
        PSR~J1803$-$6707 & run010 & $i_s$,$g_s$,$u_s$ & 59336.19351 & 162.7 & 1.532 & 1.280 & 1.75 & 110(10) & 18.8(1) & - & - \\
         & run015 & $i_s$,$g_s$,$u_s$ & 59336.42007 & 25.2 & 1.336 & 1.371 & 1.83 & 87(8) & 19.1(1) & - & - \\
         & run018 & $i_s$,$g_s$,$u_s$ & 59370.17105 & 80.0 & 1.334 & 1.272 & 1.24 & 114(7) & 18.76(7) & - & - \\
         & run008 & $i_s$,$g_s$,$u_s$ & 59371.11117 & 133.0 & 1.482 & 1.286 & 1.17 & 67(6) & 19.3(1) & - & - \\
         & run012 & $i_s$,$g_s$,$u_s$ & 59371.26172 & 88.2 & 1.269 & 1.333 & 1.18 & 115(2) & 18.75(2) & - & - \\
         & run016 & $i_s$,$g_s$,$u_s$ & 59371.39191 & 69.5 & 1.525 & 1.769 & 1.35 & 88(6) & 19.04(7) & - & - \\
         & run014 & $i_s$,$g_s$,$u_s$ & 59372.99456 & 127.7 & 2.184 & 1.571 & 1.99 & 66(5) & 19.36(8) & - & - \\
         & run026 & $i_s$,$g_s$,$u_s$ & 59406.16810 & 282.5 & 1.270 & 1.905 & 1.37 & 100(20) & 18.9(2) & - & - \\
         & run008 & $i_s$,$g_s$,$u_s$ & 59408.34457 & 32.0 & 1.804 & 1.967 & 2.46 & 73(3) & 19.24(4) & - & - \\
         & run018 & $i_s$,$g_s$,$u_s$ & 59409.04084 & 86.3 & 1.383 & 1.285 & 1.66 & 94(8) & 18.96(9) & - & - \\
         & run019 & $i_s$,$g_s$,$u_s$ & 59409.10398 & 180.5 & 1.282 & 1.352 & 1.53 & 69(4) & 19.30(7) & - & - \\
        PSR~J1823$-$3543 & run011 & $i_s$,$g_s$,$u_s$ & 59407.32350 & 19.1 & 1.539 & 1.673 & 1.59 & -11(2) & >21.74 & - & - \\
         & run023 & $i_s$,$g_s$,$u_s$ & 59409.24609 & 17.7 & 1.138 & 1.180 & 1.23 & -4(2) & >21.74 & - & - \\
        PSR~J1858$-$5422 & run031 & $i_s$,$g_s$,$u_s$ & 59406.38870 & 25.0 & 1.825 & 2.024 & 1.11 & -4(4) & >21.29 & - & - \\
         & run013 & $i_s$,$g_s$,$u_s$ & 59407.33941 & 15.7 & 1.466 & 1.532 & 1.94 & -1(2) & >21.83 & - & - \\
         & run011 & $i_s$,$g_s$,$u_s$ & 59408.37866 & 24.3 & 1.779 & 1.961 & 2.54 & 1(8) & >20.42 & - & - \\
         & run014 & $i_s$,$g_s$,$u_s$ & 59408.99809 & 28.7 & 1.641 & 1.502 & 1.74 & -3(4) & >21.09 & - & - \\
         & run031 & $i_s$,$g_s$,$u_s$ & 59409.30073 & 11.3 & 1.309 & 1.341 & 1.37 & 0(10) & >24.86 & - & - \\
        PSR~J1906$-$1754 & run024 & $i_s$,$g_s$,$u_s$ & 59406.14278 & 30.8 & 1.063 & 1.034 & 1.07 & -5(2) & >22.03 & - & - \\
         & run027 & $i_s$,$g_s$,$u_s$ & 59409.27380 & 17.0 & 1.177 & 1.228 & 1.36 & -10(20) & >24.27 & - & - \\
        \hline
    \end{tabular}
\end{table*}

\subsection{Radio discovery and timing}
\label{sec:radio_discovery}


The new pulsars were discovered during a survey of 79 unidentified \textit{Fermi}-LAT $\gamma$-ray sources by the Transients and Pulsars with MeerKAT (TRAPUM) Large Survey Project, using the 64-antenna MeerKAT radio telescope array \citep{Jonas2016+MeerKAT}. Each of the sources was observed twice for 10 minutes in L-band (856 $-$ 1712 MHz). The full technical detail of the survey and its discoveries are described in \citet{Clark2023Lband}. We found nine new MSPs; one is isolated and the remaining eight are in binaries.

Although sky positions of the pulsars can be obtained from the position of the beam containing them, it can be enhanced using triangulation of measured signal-to-noise ratio in neighbouring coherent beams. This method, named tied-array beam localization (TABLo), is implemented in a publicly available Python package, \texttt{SeeKAT}, with comprehensive details provided in \citet{Bezuidenhout2023}. As shown in Figure~\ref{fig:loc-plot}, two-sigma localisation ellipses are generally less than 10 arcseconds; small enough to contain only few optical counterparts, and most likely to be consistent with the more accurate sky positions from the follow-up timing campaign that will ensue.

Subsequent to the discovery, the TRAPUM collaboration initiated a timing campaign including both radio and $\gamma$-ray data. Comprehensive details of this campaign and its outcomes are outlined in \citet{Burgay2024+timing}. A summary of the precise positions and orbital parameters from the pulsar timing can be found in Table~\ref{tab:timing_params}. Although the lower boundary of the companion mass for these MSPs falls within the typical range for redbacks, black widows, or white dwarfs, PSR~J1823$-$3543 has an orbital period too long for a typical spider pulsar. It is more likely a white dwarf that evolved from a system with an orbital period above the bifurcation period at the onset of stable mass transfer. The other remaining binaries, however, could potentially be spider systems. We later figured out their true nature by examining optical variability in Section~\ref{sec:discussion}.

\subsection{Photometry}
\label{sec:photometry}
We observed these recently discovered pulsars to search for their optical counterparts, using ULTRACAM mounted on the New Technology Telescope (NTT) at the La Silla Observatory in Chile. The triple-beam optical design of ULTRACAM facilitates simultaneous observations across three distinct filters with high time resolution \citep{Dhillon2007}. The camera is equipped with a high-throughput version of SDSS filter set \citep{Doi2010}, denoted in this paper as $u_s$, $g_s$, $r_s$, $i_s$, and $z_s$, though all observations were performed with the $u_s$, $g_s$, $r_s$ and $i_s$ filters. In 2021, we conducted a 20-hour observation campaign on PSR~J1803$-$6707 spanning a period of three months. Additionally, we performed shorter exposures, up to 20 minutes each, for other sources than the redbacks during multiple epochs. In 2022, we also acquired an 11-hour exposure of PSR~J1036$-$4353. We routinely obtained frames of bias, dark, and twilight-sky flat images for calibration. Details of the observation are provided in Table~\ref{tab:observation_and_upper_limits}.

Images were calibrated and photometry extracted using the HiPERCAM data reduction pipeline \citep{Dhillon2018}. The images captured in $i_s$ band exhibited significant fringing artefacts. To mitigate this issue, a fringe map was created from flat fields. Aperture photometry was employed for extracting fluxes from both the target source and nearby comparison stars. The pipeline automatically adjusts the extraction aperture radius between 1.8 and 15.4 pixels (0.64–5.5\arcsec) to accommodate variations in the point-spread function (PSF) width. The maximum radius is sufficiently large to cover the worst seeing conditions of 2.54\arcsec in the dataset.

After extracting the instrumental flux using HiPERCAM, corrections were applied to account for temporal changes in air mass and transparency. The variability in transparency was mitigated using an ensemble photometry technique to obtain relative photometry \citep{Honeycutt1992}. We identified the most stable stars as comparison stars by stacking frames over 5 minutes to obtain deeper images and improve the signal-to-noise ratio for faint sources. Stellar sources on these images were detected using \texttt{photutils} \citep{Bradley2022}, and we selected a subset that presented minimal variability (RMS $\lesssim$ 0.01 mag) to build our ensemble photometry for the original single frames. We identified some segments of the PSR~J1803$-$6707 light curves that were affected by defective pixels and excluded the corresponding time window from further analysis.

Most of our fields lie beyond the sky regions covered by SDSS \citep{Abdurro'uf2022} or Pan-STARRS \citep{Flewelling2020}. To establish an absolute flux calibration, we instead cross-matched sources in the image fields to the \textit{Gaia} catalogue \citep{Prusti2016, Vallenari2023}. The mean apparent magnitudes of the corresponding \textit{Gaia} sources were converted to the equivalent ULTRACAM filter magnitudes using colour-colour relationships established by \citet{Davenport2014}, \citet{Carrasco2023} and \citet{Brown2022}. By evaluating the offsets between the relative magnitudes and the converted magnitudes of all sources in the field after removing outliers, the absolute fluxes of the targeted sources were estimated. The RMS residuals between the relative magnitudes and the converted magnitudes are below 0.01 mag in the $g_s$, $r_s$, and $i_s$ bands, and below 0.03 mag in the $u_s$ band.

We used an astrometric calibration service, \texttt{Astrometry.net}, through \texttt{Astroquery}, to obtain the coordinate transformation for the reference deep images. This astrometric data is then used to cross-match with the radio localisations of sources to identify potential optical counterparts as mentioned in Section~\ref{sec:radio_discovery}. The redback companions of PSRs~J1803$-$6707 and J1036$-$4353, were readily identified through their distinctive sinusoidal light curves. Despite the minimum mass and the orbital period of PSR~J1526$-$2744 being within the typical range of redbacks, there is no optical counterpart detected within 5 arcseconds from the radio position. None of the potential optical counterparts within the two-sigma localisation of the remaining five binary pulsars from our list show significant flux variability during and between observations. This agrees with our expectation that, given their orbital period, they should have white dwarf companions (see Table \ref{tab:timing_params}). These potential counterparts are discussed further in Section~\ref{sec:discussion}.

PSR~J1036$-$4353 displays an asymmetric double-peaked light curve with peak-to-peak amplitudes of approximately 0.35, 0.41, and 1.55 magnitudes in the $r_s$, $g_s$, and $u_s$ bands, respectively. In contrast, PSR~J1803$-$6707 has a single-peaked light curve. A comparison of its June and July 2021 data reveals a noticeable brightness difference, as shown in Figure~\ref{fig:first_lc_plot}. The peak-to-peak amplitudes in the $i_s$ band are about 0.69 mag in June and 0.74 mag in July. In the $g_s$ band, the amplitudes are 1.32 mag in June and 1.39 mag in July. For the $u_s$ band, the amplitudes are approximately 1.72 mag in June and 1.74 mag in July.

\begin{table}
    \centering
    \caption{Radial velocity of PSR~J1036$-$4353 from SOAR/Goodman spectroscopy.}
    \label{tab:j1036_radial_velocity}
    \begin{tabular}{@{\:}l|@{\:}l|@{\:}l}
    \hline
    & \multicolumn{2}{c}{Mg\textit{b}} \\ 
    BMJD & RV & $\Delta$RV\\
    & \multicolumn{2}{c}{km s$^{-1}$} \\
    \hline
    \hline
    59965.2423936   & 39.9      & 21.7 \\
    59965.2619333   & 175.7     & 19.1 \\
    59966.1517539   & 394.7     & 26.8 \\
    59966.1692315   & 247.4     & 22.7 \\
    59966.1913010   & 86.1      & 26.4 \\
    59966.2087812   & -65.3     & 16.6 \\
    59966.2310107   & -115.5    & 20.5 \\
    59966.2484887   & -125.6    & 22.0 \\
    59966.2705160   & -62.5     & 18.1 \\
    59966.3137790   & 316.4     & 20.3 \\
    59966.3312565   & 410.1     & 21.8 \\
    60004.0544388   & 354.0     & 26.5 \\
    60004.0723330   & 252.4     & 24.0 \\
    60004.0931331   & 77.0      & 23.6 \\
    60004.1113413   & -19.0     & 29.1 \\
    \hline
    \end{tabular}
\end{table}

\begin{table}
    \centering
    \caption{Radial velocity of PSR~1803$-$6707 from SOAR/Goodman spectroscopy.}
    \label{tab:j1803_radial_velocity}
    \begin{tabular}{lllll}
    \hline
    & \multicolumn{2}{c}{Metal lines} & \multicolumn{2}{c}{Mg\textit{b}} \\ 
    BMJD & RV & $\Delta$RV & RV & $\Delta$RV \\
    & \multicolumn{2}{c}{(km s$^{-1}$)} & \multicolumn{2}{c}{(km s$^{-1}$)} \\
    \hline
    \hline
    59306.3604777   & -187.3    & 32.5   & -152.4   & 36.9  \\
    59348.2622252   & -122.8    & 35.2   & -106.1   & 36.3  \\
    59348.2801643   & -57.6     & 18.4   & -78.7    & 22.0  \\
    59375.2975885   & -14.7     & 27.4   & -23.6    & 29.3  \\
    59375.3159278   & 47.3      & 21.3   & 33.1     & 22.9  \\
    59375.3390443   & 150.6     & 24.4   & 142.2    & 31.3  \\
    59375.3585664   & 234.7     & 45.6   & 258.2    & 47.8  \\
    59375.3906485   & 300.5     & 24.7   & 277.2    & 27.9  \\
    59375.4085084   & 294.8     & 31.4   & 294.2    & 33.4  \\
    59398.3164252   & 109.1     & 34.7   & -        & -     \\
    59411.1657053   & 261.6     & 32.3   & 247.2    & 33.6  \\
    59411.1834266   & 312.8     & 17.9   & 301.7    & 22.7  \\
    59411.2207368   & 196.2     & 19.4   & 261.6    & 26.7  \\
    59411.2488346   & 127.6     & 16.2   & 126.7    & 18.7  \\
    59411.2667791   & 73.9      & 25.3   & 63.7     & 29.7  \\
    59411.2908973   & 25.5      & 26.7   & 27.7     & 28.5  \\
    59411.3087203   & -73.9     & 17.2   & -74.8    & 18.1  \\
    59424.3111661   & -198.6    & 25.7   & -156.1   & 34.7  \\
    59424.3290357   & -186.8    & 31.6   & -139.5   & 35.4  \\
    59426.3077428   & 78.7      & 26.8   & 27.1     & 38.8  \\
    59426.3252472   & 139.8     & 35.8   & 131.5    & 45.4  \\
    \hline
    \end{tabular}
\end{table}

\subsection{SOAR/Goodman Spectroscopy}

Optical spectroscopy for the two redback targets was obtained using the red camera of the Goodman Spectrograph \citep{Clemens2004} on the 4.1m SOAR telescope in Chile. The spectroscopic data of PSRs~J1803$-$6707 and J1036$-$4353 were obtained in mid 2021 (21 spectra over seven nights) and early 2023 (15 spectra over three nights), respectively. AB magnitudes of both redbacks are $>$18. Hence, individual exposure times were set to 25 minutes per spectrum. All spectra used a 400 line mm$^{-1}$ grating and either a 0.95\arcsec or 1.2$\arcsec$ slit, giving a FWHM resolution of $\sim 5.6$ \AA\ (0.95\arcsec) or $\sim 6.7$ \AA\ (1.2\arcsec) over a useable wavelength range of $\sim$4000--7800 \AA.

All spectra were reduced and optimally extracted using \texttt{IRAF} \citep{Tody1986,Tody1993,Fitzpatrick2024}. Our methodology for wavelength calibration and radial velocity measurement is described in detail by \citet{Dodge2024}, updating the methodology used by \citet{Strader2019}. In brief, initial wavelength calibration is done using FeAr arc lamp exposures taken adjacent to each spectrum. Zero-point corrections, which account for flexure or slit miscentering, are calculated by fitting telluric absorption bands in each spectrum using models calculated via {\tt TelFit} \citep{Gullikson2014}. We then determined radial velocities fitting over a grid of PHOENIX \citep{Allard2016} templates convolved to the appropriate SOAR resolution and allowing for rotational broadening, using the package {\tt RVSpecFit} \citep{Koposov2011,Koposov2019}. For the J1036$-$4353 spectrum, we only fit around the Mg$b$ region, which provides metal-line absorption velocities that are less sensitive to irradiation \citep{Linares2018}. For J1803$-$6707, the S/N in the Mg$b$ region alone was not sufficiently high for reliable measurements, so we instead fit simultaneously around Mg$b$ and the metal lines in the range 6050--6375 \AA. For comparison, we also fit the full spectral region  (4000--6800 \AA). The results are shown in Table~\ref{tab:j1036_radial_velocity} and \ref{tab:j1803_radial_velocity}.

\begin{figure}
    \centering
    \includegraphics[width=\columnwidth]{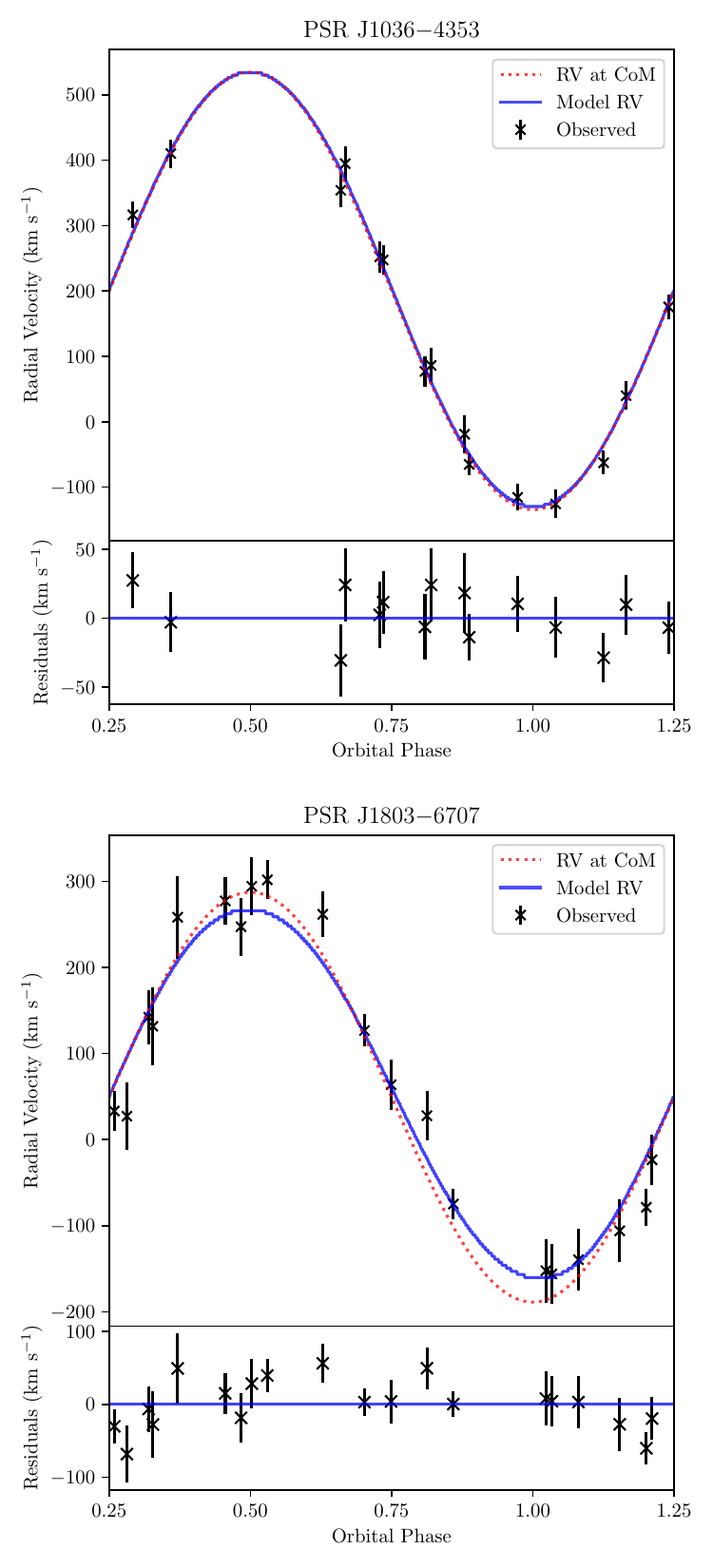}
    \caption{The radial velocity curves of PSRs~J1036$-$4353 and J1803$-$6707 are shown as crosses with vertical bars of uncertainties. The blue solid lines are the results of cross-correlating model spectra with a template (defined as the model at orbital phase = 0.25) to reproduce the measured radial velocities. The red dotted lines are the sine curves which are approximately radial velocities tracing the centre of mass of the companions. The discrepancy between the model of
    PSR~J1803$-$6707 and the sine curve shows the clear effect of how radial velocity is shifted from the centre of mass to the `centre of light' toward the brighter, hotter region of the star.}
    \label{fig:rv_plot_j1036_and_j1803}
\end{figure}

\section{Modelling}
\label{sec:modelling}

We modelled the optical light and radial velocity curves of PSRs~J1036$-$4353 and J1803$-$6707 using the stellar binary light curve synthesis code, \texttt{Icarus} \citep{Breton2011}. In this code, the companion and its orbit are parameterised by the mass ratio ($q$), orbital period ($P_{\rm orb}$), orbital inclination ($i$), radial velocity amplitude ($K_2$), ratio of rotational to orbital speeds ($\omega$), filling factor ($f_{\rm RL}$), exponent of the gravity darkening relation ($\beta$), base temperature ($T_{\rm base}$), and irradiation temperature ($T_{\rm irr}$). $q$ is defined as the ratio of the pulsar mass to the companion mass. The companion is formed as a tessellated surface of four-sided facets using a healpix projection \citep{Gorski2005}. After receiving the parameters, the equipotential surface, temperatures, and gravity darkening are calculated for different points on the grid, followed by an evaluation of the orbital separation and stellar masses. Then, model fluxes are evaluated using the atmosphere grid created from the atmosphere library \texttt{ATLAS9} \citep{Castelli2004}. Full detail on implementing the library in our code can be found in \citet{Kennedy2022}. To fit the observed data, two additional free parameters are included: total extinction in V band ($A_V$) and the distance ($d$). The code has been widely employed to infer orbital and stellar parameters in spider pulsars \citep[e.g][]{Sanchez2017, Voisin2020, Kennedy2022, MataSanchez2023, Dodge2024}. It was also extended to handle asymmetric light curves \citep{Voisin2020} and spectroscopic data \citep{Kennedy2022, Dodge2024}. In our case, the availability of a precise pulsar ephemeris means that $P_{\rm orb}$ can be fixed, and $q$ does not need to be fitted for, but is instead derived from the pulsar mass function in conjunction with the input $K_2$ and $i$. As is usually done, $i$ is assumed to be uniformly distributed in $\cos i$ (i.e. assuming random orbital inclinations).

There are several explanations to the observed asymmetry in the light curves. \citet{Romani2016} proposed that the asymmetric irradiation is from non-thermal X-ray emission generated in the intra-binary shock (IBS) between the pulsar and companion winds. However, \citet{Zilles2020}, using particle shower simulation, subsequently demonstrated that the irradiation from X-ray emission alone is inadequate to explain the asymmetry. Alternatively, \citet{Kandel_2020_heated_pole} suggested that the observed asymmetry is due to hot spots heated by high-energy particles ducted down to the magnetic poles. \citet{Voisin2020} introduced heat redistribution models to enhance fitting for systems exhibiting asymmetric light curves. They superficially modelled heat transport over stellar surface using the surface parallel energy transport equation. This equation includes two terms: one for diffusion and the other for convection. The convection term can be prescribed to reflect different theoretical models. Subsequently, \citet{Stringer2021} incorporated a power law dependency on the temperature into the diffusion term.

Most redback companions are affected by irradiation. In some cases this leads to a large temperature contrast between the day and night sides; in other cases, the effects of irradiation are subtle, but can be revealed by careful light curve modelling. Other factors including heat transport, ellipsoidal distortion, and gravity darkening also change temperature gradient across the stars. This results in variation of spectral line strength across different parts of the companions. As a result, the radial velocity from a spectral line will deviate from the radial velocity at the centre of mass (CoM) of the companion differently along the orbit \citep[see, e.g.][]{Wade1988,vanKerkwijk2011}. The observed radial velocity is then called the centre of light (CoL). Obtaining accurate radial velocity correction between the CoL and CoM is crucial for accurate estimation of pulsar and companion mass. Self-consistent spectral modelling inherently accounts for CoL corrections when modelling the radial velocity curve. \citet{Kennedy2022} performed a precise neutron star mass measurement by using whole-spectral modelling to enhance radial velocity fitting. Similarly, \citet{Dodge2024} focused on a specific spectral range around the Mg\textit{b} triplet to improve computational speed, while \citet{Linares2018} used the equivalent width as a proxy to constrain the average line emission location. In this work, the radial velocity curve is also obtained by simulating spectra around the Mg\textit{b} triplet and measuring the radial velocity on the simulated line, as implemented in \citet{Dodge2024}, such that the fitted $K_2$ can self-consistently incorporate the effects of irradiation. The $K_{2}$ correction between CoL and CoM is particularly important for stars with large temperature contrasts, such as PSR~J1803$-$6707 while it is minimal in the case of PSR~J1036$-$4353 as shown in Figure~\ref{fig:rv_plot_j1036_and_j1803}.

High-energy particles and $\gamma$-ray can deposit heat below the photosphere leading to higher effective temperature on the companion \citep[see, e.g.,][]{Zilles2020}. It is found that strong irradiation changes the surface boundary conditions of hot Jupiters that reduce cooling rate of the planets \citep{Guillot1995, Arras2006}. \citet{Ginzburg2020} showed that the suppression of cooling rate of the companions due to strong irradiation plays a role in the evolution of spider pulsar binaries. \citet{Romani2021} applied gravity darkening after irradiation for mass estimation for PSR~J1810+1744. \citet{Dodge2024} took this concept and compared the two different prescriptions of gravity darkening and irradiation processes. The first prescription, pre-irradiation gravity darkening (Pre-IGD), is the same as earlier uses of \texttt{Icarus} (and adopted more widely in the field so far) in that the gravity darkening is applied before the irradiation and heat redistribution. This is expected to replicate the effects after shallow heating. In the second prescription, post-irradiation gravity darkening (Post-IGD), the gravity darkening is calculated after the irradiation effects to replicate the effects of heat deposition deeper below the photosphere.

We fitted the redbacks with the direct heating (DH) model in which incident radiations are promptly absorbed, reprocessed, and emitted \citep{Breton2013}. The optical light curve of PSR~J1036$-$4353 is clearly asymmetric as the first peak is higher than the second, as shown in Figure~\ref{fig:first_lc_plot}. Thus, it is fitted with stellar spot, and heat redistribution models. The latter is the same model used by \citet{Voisin2020,Stringer2021,Dodge2024}. In these models, we used the convection profile with constant angular velocity (around the spin axis of the star) from \citet{Voisin2020}. The profile assumed a superficial layer of constant thickness, density, and thermal capacity across the entire stellar surface. In the diffusion term, there are two parameters in this model, the diffusion coefficient ($\kappa$) and the diffusion index ($\Gamma$) \citep[see][for more details]{Stringer2021}. We also separated these models into three configurations starting from fixing both parameters, freeing $\kappa$, and freeing both parameters. These are convection-only (C), diffusion and convection (D+C), and power-law diffusion and convection (PD+C) models.

PSR~J1803$-$6707 has noticeable discrepancies between the light curves over the observation time span as shown in Figure~\ref{fig:first_lc_plot}. For this reason we separated the data into three sets according to the observing epoch they were obtained. In this configuration, the joint modelling incorporated a subset of parameters that share common values between the three epochs and others which they are allowed to vary between them. The group of parameters varying from epoch to epoch included base temperature, irradiation temperature and Roche-Lobe filling factor. Because the first epoch of observing in May 2021 covers less than half of an orbital cycle, we decided to only model the data from the latter two epochs (June and July 2021).

\begin{figure}
    \centering
    \includegraphics[width=\columnwidth]{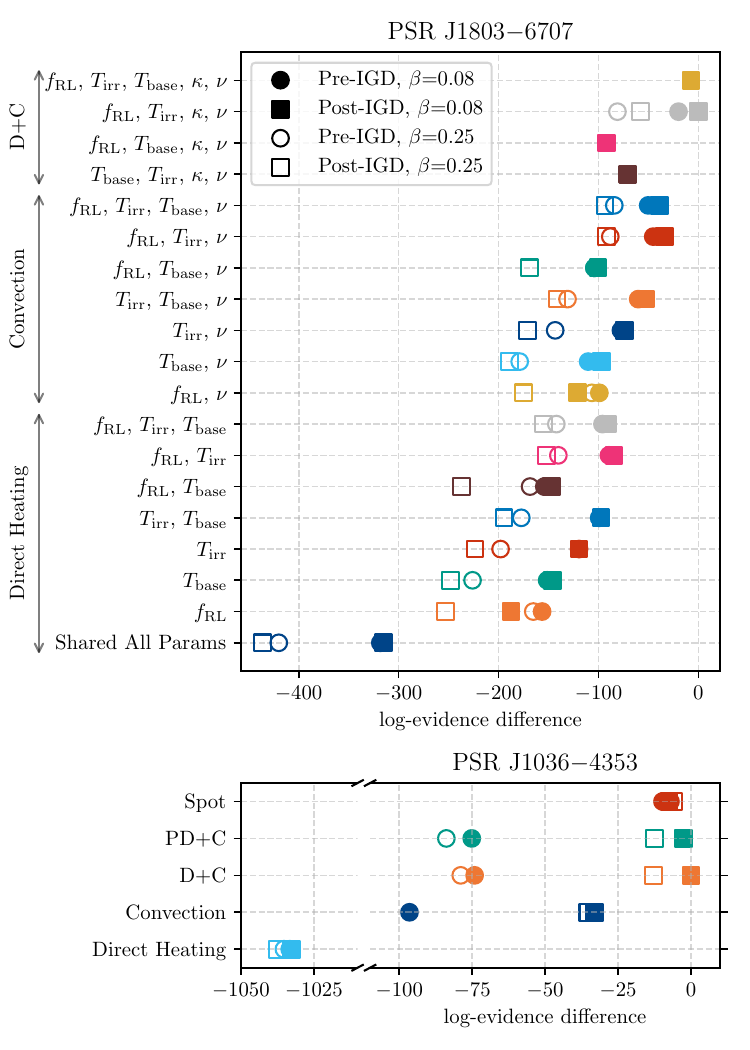}
    \caption{Different models are compared using their Bayesian evidence relative to the best-fitting model. As described in Section~\ref{sec:modelling}, a decisively better model will have a log-evidence higher by 4.6 compared to another model. The log-evidence values are labelled according to the set of parameters that were separately fitted to different datasets. This implies that the D+C model with Post-IGD are the best-fitting models for both PSR~J1036$-$4353 and PSR~J1803$-$6707. Specifically, the D+C model with Post-IGD and separate $f_{\rm RL}$, $T_{\rm irr}$, $\kappa$, and $\nu$ parameters for different datasets best describes the modulation changes in PSR~J1803$-$6707.}
    \label{fig:model_comparison}
\end{figure}

Radio timing results from \citet{Burgay2024+timing} including the projected semi-major axis (A1), the binary orbital period (P$_{\rm b}$), the epoch of ascending node (T$_{\rm asc}$), the barycentric rotation frequency (F0), and the time derivative of the barycentric rotation frequency (F1) are used for parameter derivations during the optical modelling. The pulsar position, proper motion, and annual parallax were taken from \textit{Gaia} DR3 catalogue \citep{Vallenari2023}. The prior distributions for colour excess E(B$-$V), associated with reddening in the line of sight to the object, are assumed to be Gaussian distributions. We used E(B$-$V) of 0.1411 and 0.0665 for PSRs~J1036$-$4353 and J1803$-$6707, respectively. These values are from \citet{Chiang2023} which is based on \citet{Schlegel1998}. We set the standard deviation of the prior to be 0.03, confining the colour excess within a physically realistic range around the provided mean values. The \textit{Gaia} parallaxes are 0.36(33)~mas and 0.17(27)~mas for PSRs J1036$-$4353 and J1803$-$6707, respectively \citep{Vallenari2023}. The prior of the distance to the binaries is estimated from the joint probability of the Galactic distribution of MSPs \citep{Levin2013}, the parallax, and the distance estimated from radio dispersion measurement (DM) using the electron density model from \citet[][hereafter YMW16]{YMW2016}. \citet{Espinosa2012} showed that gravity darkening exponents are not significantly different for stars with mass ratios similar to those of black widow and redback systems. Hence, $\beta$ is expected to be in the range between 0.23 and 0.25. However, their model did not account for the effect of irradiation. For this reason we also fit the data with both $\beta$=0.25 and 0.08, with the latter the typical value for fully convective stars \citep{Lucy1967}.

The availability of photometry data in multiple bands is crucial in constraining the temperatures and colour excess. Small differences in instrumental filters and calibration standards alongside with uncertainties in atmosphere models are likely to cause systematic offsets between photometric bands that will result biased colours. For this reason we allow offsets in the predicted light curves compared to our observations. These are subjected to Gaussian priors centred a zero offset, and with standard deviations of 0.06 mag in the $u_s$ band and 0.02 mag in the other bands. These standard deviations are twice the RMS of the residuals between the model and the observations. We believe these reflect the degree of uncertainty there is in the calibration, with $u_s$ being notoriously poorer due to the smaller number of reference stars available to establish the instrumental zero point. We also tightened these priors for a subset of models, but the goodness-of-fit ranking was unchanged.

The modelling was optimised under Bayesian inference using \texttt{PyMultiNest} \citep{Buchner2014} which is the Python wrapper to \texttt{Multinest} library \citep{Feroz2009}, an implementation of the nested sampling algorithm \citep{Skilling2004}. The aim of this algorithm is to estimate the Bayesian evidence ($Z$) which is useful to compare models. Given the wide variety of options available to us in terms of irradiation prescriptions and selection of parameters which can be fixed or fitted between epochs, we performed a large number of model inference. When comparing two models, $M_1$ and $M_2$, the ratio of the evidence values, $Z_{M_1}/Z_{M_2}$, also known as Bayes factor ($B_{1,2}$), is usually calculated and assessed using Jeffreys' scale \citep{Jeffreys1939}. The scale suggests that $B_{1,2}>100$ decisively favours $M_1$ over $M_2$ while $1 < B_{1,2} < 10^{1/2}$ is not significant (and should prioritise the simpler model as per Occam's razor). In this paper, we use the natural logarithmic values of the evidences and Bayes factors, i.e. $\ln{(B_{1,2})} = \ln{(Z_{M_1})} - \ln{(Z_{M_2})}$. The above decision boundaries therefore translate to approximately $\ln{(B_{1,2})} = 4.6$ and 1.2, respectively. Figure~\ref{fig:model_comparison} shows the log-evidence relative to the (best) maximum log-evidence for the variations of models we have tested. Table~\ref{tab:j1036_and_j1803_params} presents full posterior parameters and derived quantities for the model that best fits each of the two redbacks.

\begin{table}
    \centering
    \renewcommand{\arraystretch}{1.8}
    \caption{Fitted and derived parameters of best-fitting models which are the diffusion and convection models with post-irradiation gravity darkening prescription for both PSR~J1036$-$4353 and PSR~J1803$-$6707. These models have the gravity darkening exponent $\beta=$0.08. $\chi^{2}_{\rm \nu}$, the reduced chi-squared value, is showed as an indication of goodness-of-fit. $v \sin{(i)}$ is the projected rotational velocity of the companion. $\varepsilon$ is the irradiation efficiency, defined as the ratio of the irradiation energy rate to the spin-down luminosity. $T_{\rm day}$ and $T_{\rm night}$ are the average temperatures of the companion's hemispheres facing toward and away from the pulsar, respectively. $T_{\rm dusk}$ and $T_{\rm dawn}$ are the average temperatures of the eastern and western hemispheres of the companion, respectively.}
    \label{tab:j1036_and_j1803_params}
    \begin{tabular}{l|c|c|c}
        \hline
        Source & PSR~J1036$-$4353 & \multicolumn{2}{c}{PSR~J1803$-$6707} \\
        \hline
        Model & D+C & \multicolumn{2}{c}{D+C} \\
        \hline
        Dataset & All & June 2021 & July 2021 \\
        \hline
        \multicolumn{4}{c}{Fitted Parameters} \\
        E(B$-$V) & $0.14\pm0.02$ & \multicolumn{2}{c}{$0.008_{-0.006}^{+0.01}$} \\
        $K_{\rm c}$ (km s$^{-1}$) & $340_{-10}^{+9}$ & \multicolumn{2}{c}{$244_{-10}^{+9}$} \\
        $d$ (kpc) & $3.4\pm0.1$ & \multicolumn{2}{c}{$3.3\pm0.1$} \\
        $i$ ($^\circ$) & $86_{-4}^{+2}$ & \multicolumn{2}{c}{$53\pm2$} \\
        $T_{\rm base}$ (K) & $5380_{-100}^{+90}$ & \multicolumn{2}{c}{$4250_{-10}^{+20}$} \\
        $T_{\rm irr}$ (K) & $5100_{-300}^{+600}$ & $4740\pm40$ & $4670_{-30}^{+40}$ \\
        $f_{\rm RL}$ & $0.812_{-0.005}^{+0.006}$ & $0.772\pm0.006$ & $0.834\pm0.006$ \\
        $\nu$ (W K$^{-1}$ m$^{-2}$) & $90000_{-20000}^{+50000}$ & $-770\pm90$ & $570\pm60$ \\
        $\kappa$ (W K$^{-1}$ m$^{-2}$) & $40000_{-10000}^{+30000}$ & $1500\pm100$ & $20_{-20}^{+30}$ \\
        \multicolumn{4}{c}{Derived Parameters} \\
        $q$ & $6.1\pm0.2$ & \multicolumn{2}{c}{$4.0\pm0.2$} \\
        $M_{\rm p}$ (M$_\odot$) & $1.4\pm0.1$ & \multicolumn{2}{c}{$1.7\pm0.2$} \\
        $M_{\rm c}$ (M$_\odot$) & $0.24\pm0.01$ & \multicolumn{2}{c}{$0.44_{-0.04}^{+0.05}$} \\
        $T_{\rm dusk}$ (K) & $5570\pm70$ & $4440_{-10}^{+20}$ & $4420_{-10}^{+20}$ \\
        $T_{\rm day}$ (K) & $5590\pm70$ & $4900\pm20$ & $4890\pm20$ \\
        $T_{\rm dawn}$ (K) & $5460\pm60$ & $4460_{-10}^{+20}$ & $4400_{-10}^{+20}$ \\
        $T_{\rm night}$ (K) & $5460\pm60$ & $4160_{-10}^{+20}$ & $4140_{-10}^{+20}$ \\
        $R_{\rm c}$ (R$_\odot$) & $0.456\pm0.008$ & $0.70\pm0.02$ & $0.73\pm0.02$ \\
        $v \sin{(i)}$ (km s$^{-1}$) & $89\pm1$ & $75\pm1$ & $78\pm1$ \\
        $f_{\rm VA}$ & $0.945\pm0.003$ & $0.918\pm0.004$ & $0.956\pm0.003$ \\
        $\varepsilon$ & $0.24_{-0.06}^{+0.1}$ & $0.21\pm0.02$ & $0.20\pm0.02$ \\
        \multicolumn{4}{c}{Goodness of fit} \\
        $\ln{\rm (Z)}$           & -2267.6 & \multicolumn{2}{c}{-6622.2} \\
        N$_{\rm dof}$   & 3718 & \multicolumn{2}{c}{12485} \\
        $\chi^{2}_{\rm \nu}$ & 1.247 & \multicolumn{2}{c}{1.053} \\
        \hline
    \end{tabular}
\end{table}

\section{Discussions}
\label{sec:discussion}

\begin{figure}
    \centering
    \includegraphics[width=\columnwidth]{plots/hr_diagram_ultracam_distance_limits_all.png}
    \caption{The Hertzsprung--Russell diagram displays PSR~J1036$-$4353, PSR~J1803$-$6707, and upper limits for other pulsars. The vertical axis denotes the absolute magnitude $M_{\rm G}$ in the {\it Gaia} G band, while the horizontal axis represents the {\it Gaia} colour ($G_{\rm BP}-G_{\rm RP}$). Black dots denote {\it Gaia} sources within a 200 pc distance, with the orange circle marking the Sun's position. Magnitudes are uncorrected for interstellar extinction. The black arrow shows the shift from the intrinsic to the observed position of a star with E(B$-$V) $\simeq 0.3$ mag. ULTRACAM $i_s$ and $g_s$ magnitudes at the signal-to-noise ratio of unity are approximately 25.253 and 26.256, respectively \citep{Dhillon2001}. These limits are converted to {\it Gaia} magnitudes, establishing 1-sigma upper limits at various distances, depicted as grey dashed lines. One of these lines serves as an approximate boundary, distinguishing detectable sources above it from undetectable ones below. Upper limits for other TRAPUM pulsars are estimated from DM distance \citep[YMW16;][]{YMW2016} and represented as dashed lines in different colours.}
    \label{fig:ultracam-limit-on-gaia-hr}
\end{figure}

\subsection{Discovery and Potential Companions}

We identified the two optical counterparts of PSRs~J1036$-$4353 and J1803$-$6707, out of 8 MSP binaries discovered by the TRAPUM collaboration \citep{Clark2023Lband}. A few stars appear within the two-sigma localisation ellipses of the other sources as shown in Figure~\ref{fig:loc-plot}. All of the potential candidates apart from the redbacks do not show significant variability neither during an observation nor between epochs.

The limiting magnitudes of ULTRACAM five-minute exposure images are about 25 and 26 in $i_s$ and $g_s$ bands, respectively \citep{Dhillon2001}\footnote{\url{http://www.vikdhillon.staff.shef.ac.uk/ultracam/sensitivity.html}}. MSP companions with an absolute magnitude $\gtrsim 9$ typical of white dwarfs would not have been detected in our observations if they lied beyond $\sim$1 kpc. We can see from Figure~\ref{fig:ultracam-limit-on-gaia-hr} that the two redbacks PSRs~J1036$-$4353 and J1803$-$6707 are located 2-3 mag fainter than the Sun on the blue side of the main sequence. They were well detected at distances of 3.4 and 3.3 kpc, respectively, as inferred from optical modelling. We used the DM distances to estimate the lower limit on the absolute magnitude for the other binaries. Except for PSR~J1906$-$1754, these lower limits range from slightly to substantially below the redbacks. In particular, the apparent sources in the images of PSRs J1623$-$6936 and J1858$-$5422 are unlikely to be their optical counterparts if they are white dwarfs or faint black widows.

The orbital periods of PSRs J1623$-$6936, J1757$-$6032, J1823$-$3543, J1858$-$5422, and J1906$-$1754 fall well outside the typical range for black widows and redbacks, suggesting they are most certainly MSP binaries with white dwarf companions. Their minimum masses indicate that the white dwarfs probably contain a degenerate He core, except for PSR~J1757$-$6032 which lies near the boundary between He- and CO-core white dwarfs. This therefore confirms our expectation that none of the nearby stars are the counterpart of their companions.

At an orbital period of 4.8 hours, PSR~J1526$-$2744 falls within the range observed in spiders. However, if it were a redback similar to other detected in this paper at the DM-inferred distance of 1.3 kpc, we would have detected it. Its real distance would need to be at least $\sim$3.5 kpc to avoid detection. Given that this is probably unlikely the case, an alternative scenario is that the system is a black widow, where the companion is on average fainter. This possibility is, however, disfavoured by the minimum companion mass derived from the timing (assuming a $1.4\,{\rm M}_\odot$ pulsar) of $0.08\,{\rm M}_\odot$. This conclusion is further reinforced by the fact that our optical data includes the superior conjunction of the companion where it would show its heated side and could have been detected. We therefore conclude that the most likely nature of this system is a MSP with a white dwarf companion in a very compact orbit, unless the YMW16 distance is severely underestimated. These extremely low-mass white dwarfs are called ELM WDs system \citep[see, e.g.][for more details]{Marsh1995}. While unusual, a handful of these systems containing an MSP have been discovered, including PSR~J1012+5307 which recent optical work showed it possibly contains the lowest mass white dwarf in an MSP binary found so far at $0.15 \pm 0.02\,{\rm M}_\odot$ \citep{Mata2020}. For PSR~J1526$-$2744's companion to have a similar mass would imply the orbit is seen relatively face on ($\lesssim 40^\circ$). A white dwarf companion would also explain the lack of radio eclipses commonly seen in spider systems.

\begin{figure*}
    \centering
    \includegraphics[width=\textwidth]{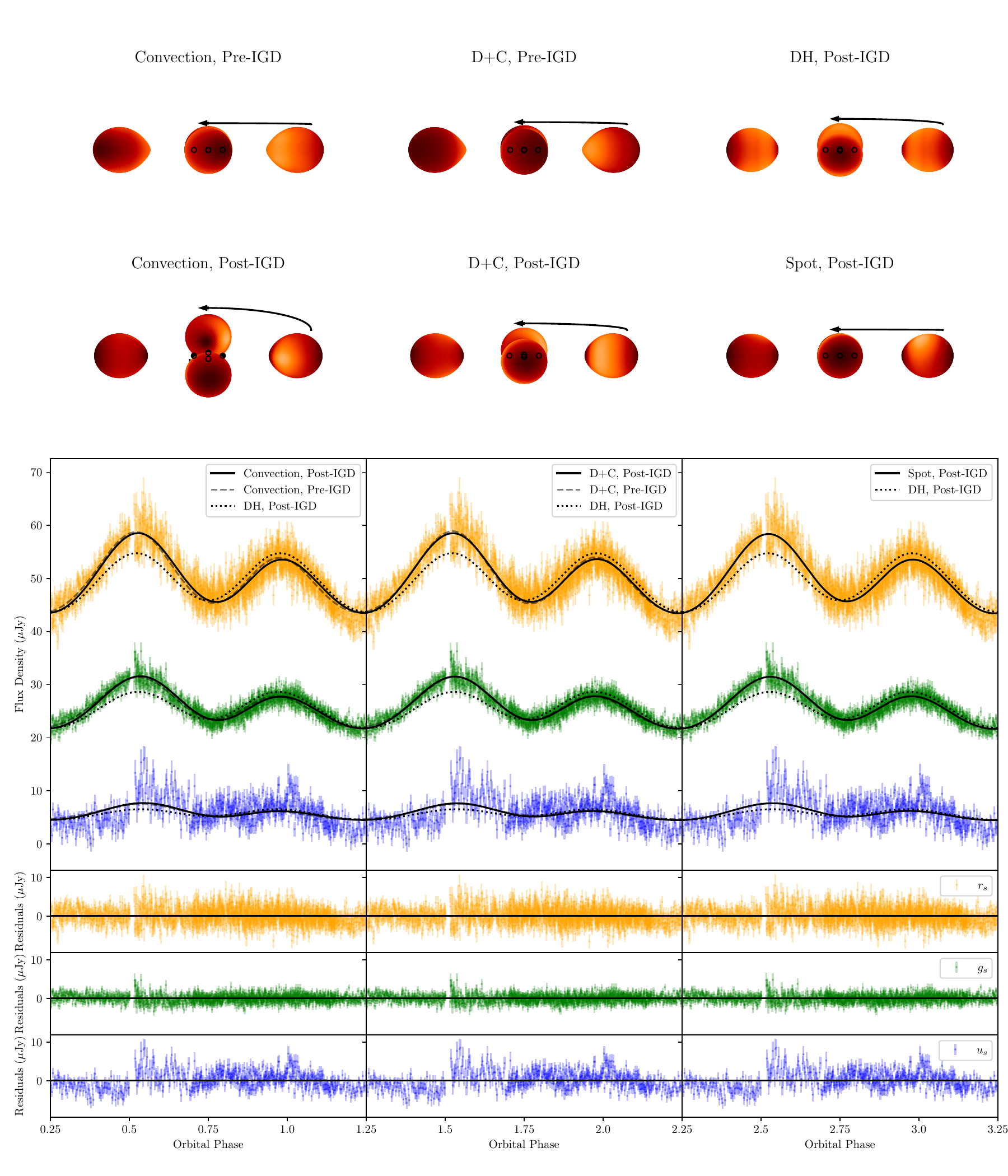}
    \caption{\textit{Top panels}: Surface temperature profiles of PSR~J1036$-$4354's companion for six different models: pre-IGD direct heating, post-IGD spot, pre-IGD convection, post-IGD convection, pre-IGD D+C (diffusion and convection), and post-IGD D+C models. \textit{Bottom panels}: Light curves of PSR~J1036$-$4354 for various model predictions. The light curve of the direct heating model with the post-IGD prescription is shown with a dotted line in all three columns. The light curves of the convection model with a uniform angular convection profile are displayed in the leftmost column. The results of the diffusion and convection model with a uniform angular convection profile are presented in the middle column. In these two columns, the grey dashed lines represent models with the pre-IGD prescription, while the black solid lines indicate the post-IGD models. The spot model is shown in the rightmost column. The light curves of the convection model with the pre-IGD prescription exhibit a phase offset compared to the observed data.}
    \label{fig:lc_1036}
\end{figure*}

\begin{figure*}
    \centering
    \includegraphics[width=\textwidth]{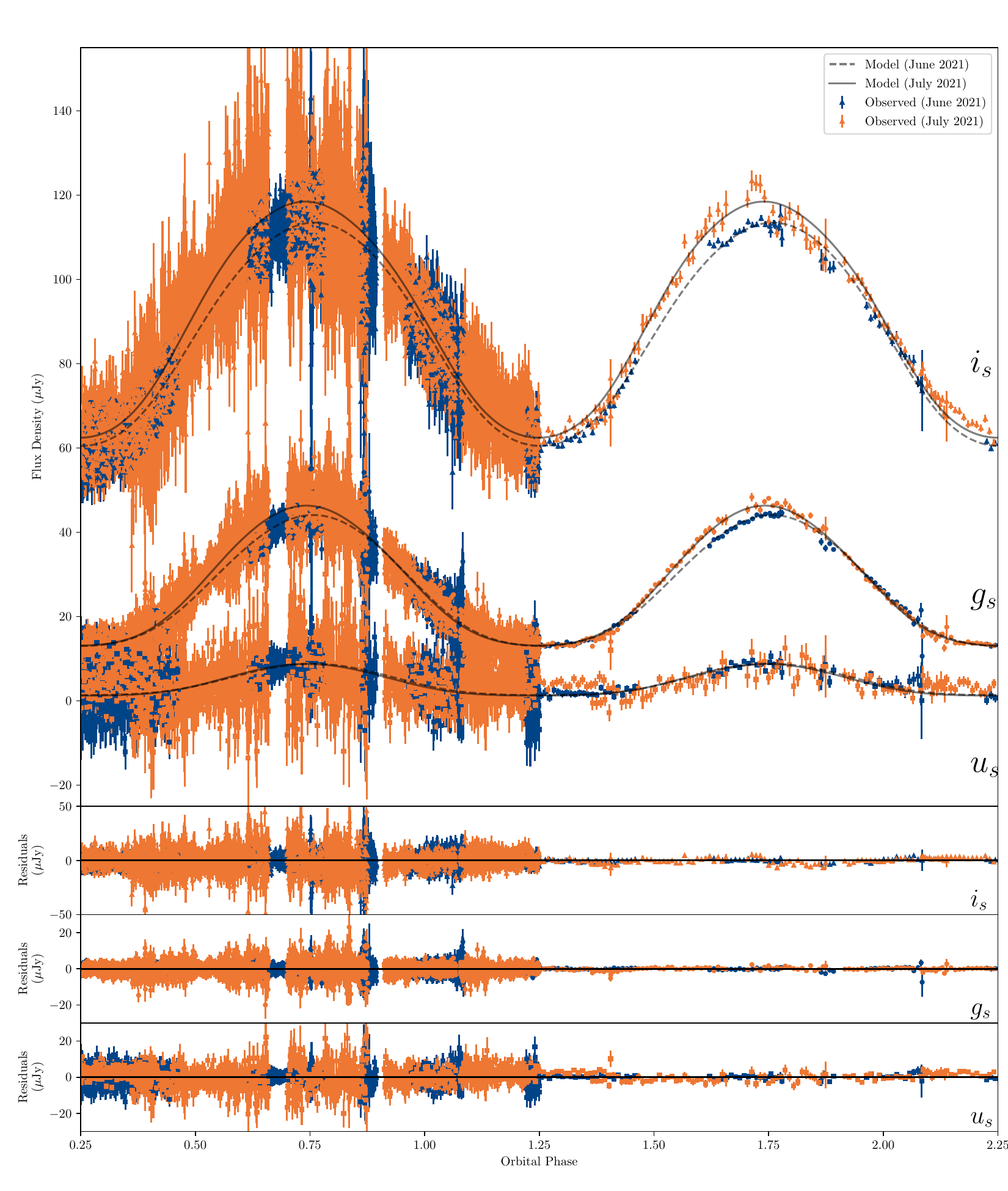}
    \caption{Light curves of PSR~J1803$-$6707 presented for two orbital cycles. The original data points are shown for the first orbital cycle (phase 0 to 1). For the second orbital cycle (phase 1 to 2), data points are binned to five-minute exposures. Markers with blue and orange colours represent data from June and July 2021, respectively. The black dashed and solid lines show the best-fitting model lines for June and July 2021, respectively. The best-fitting model is the diffusion-convection model with the post-IGD prescription and $\beta$=0.08.}
    \label{fig:lc_1803}
\end{figure*}

\subsection{Heat Deposition Below Photosphere}

PSR~J1036$-$4353 displays an asymmetric double-peak light curve whose peak at orbital phase 0.25 is higher than the other peak at orbital phase 0.75 (see Figure~\ref{fig:first_lc_plot}). The direct heating model cannot reproduce this asymmetric feature. Consequently, we proceeded to fit the data using the heat redistribution and spot models. Although the diffusion-convection model returned the best-fit statistics, its Bayesian evidence value does not differ substantially from the spot model. The inferred inclination from our modelling is about 86$^{\circ}$, which is also consistent with the extended radio eclipses (orbital phase range 0.05 to 0.45) reported by \citet{Burgay2024+timing}. At this inclination, we might also expect $\gamma$-ray eclipses to occur as the pulsar passes behind its companion at superior conjunction \citep{Clark2023eclipse}. Unfortunately, $\gamma$-ray pulsations have not been detected from this system, owing to its faintness, and so the orbital period cannot be determined from gamma-ray timing to achieve sufficient precision for detecting a $\gamma$-ray eclipse.

The fitted and derived parameters from the best-fitting model are presented in Table~\ref{tab:j1036_and_j1803_params}. The projected radial velocity amplitude of the companion of 340$^{+9}_{-10}$ km s$^{-1}$ combined with the orbital inclination from above implies a pulsar mass 1.4$\pm0.1$ M$_{\odot}$ and a companion mass 0.24$\pm0.01$ M$_{\odot}$. The day and night side\footnote{As evaluated from a viewpoint along the orbital plane.} temperatures are respectively 5590 and 5460 K while the temperatures on the `dusk' and `dawn' side are respectively 5570 and 5460 K. With our best-fit model requiring a free diffusion parameter, more heat can be transported in  the north-south direction. This means the model requires more irradiation, resulting in a hotter day side than comparable non-diffusion models such as the simple direct heating. This leads to a larger Roche-lobe filling, a higher orbital inclination, and smaller component masses (see Table \ref{tab:j1036_parameter_table} for more details).

As shown in Figure~\ref{fig:lc_1036}, the modelled light curves in which gravity darkening is applied before thermal transport (i.e. Pre-IGD) a small phase offset compared to the observed light curves, indicating a shift in the outermost layer of the star. The lower temperature regions at the `nose' and `back' points of the star move to the east, leading to phase shifts in the light curves. These displaced lower temperature regions result in larger filling factors, higher inclination, and higher irradiation due to enhanced heat transport. In response to this change, the outermost layer is further shifted away from both points. This discrepancy suggests that the heat redistribution models with shallow heating fail to reproduce the observed light curves. In addition, the Bayesian evidence for these models with Pre-IGD is notably lower than their counterparts with Post-IGD, as indicated in Figure~\ref{fig:model_comparison}.

The Bayesian evidence values for the heat redistribution models with gravity darkening applied after thermal transports (i.e. Post-IGD) show a significant improvement ($\ln{(B)} > 60$) in their fit to the PSR~J1036$-$4353 data. The Post-IGD keeps the temperature pattern from gravity darkening in place and hence requires lower irradiation and surface wind speed. This means that most of the irradiation energy can be deposited deeper than the photosphere before being redistributed to other parts of the companion surface. This agreed with the assumption in \citet{Zilles2020} and \citet{Ginzburg2020} that pulsar irradiation penetrates deeper than the photosphere and raises the effective temperature of the companion. This deep energy deposition leads to the formation of a deep radiative layer in the convective envelope, slowing down the companion's cooling. The direct heating and spot models do not present significant differences in their goodness-of-fit between the Pre- and Post-IGD alternatives since these irradiation prescriptions do not involve energy transport across the companion.

The phase shifts found with the Pre-IGD prescriptions are not suitable to model the light curves of PSR~J1036$-$4353. On the other hand, the Post-IGD prescriptions do not create this effect, thus supporting the deep penetration theory of high-energy photons. A recent study of the asymmetric, irradiation-dominated, single-peak light curves of PSR~J1910$-$5320 \citep{Dodge2024} found mixed results as to whether Pre- or Post-IGD models performed better, with a slight advantage in favour of the latter. However, their data were fitted with a gravity darkening $\beta$=0.25 while we compared both $\beta$=0.25 and $\beta$=0.08 in this study and found that models using the latter outperformed the former. It is important to note that current gravity darkening models do not account for the effects of irradiation and magnetic fields, making it challenging to ascertain if Post-IGD is the universally applicable prescription for all spider binaries.

\begin{figure}
    \centering
    \includegraphics[width=\columnwidth]{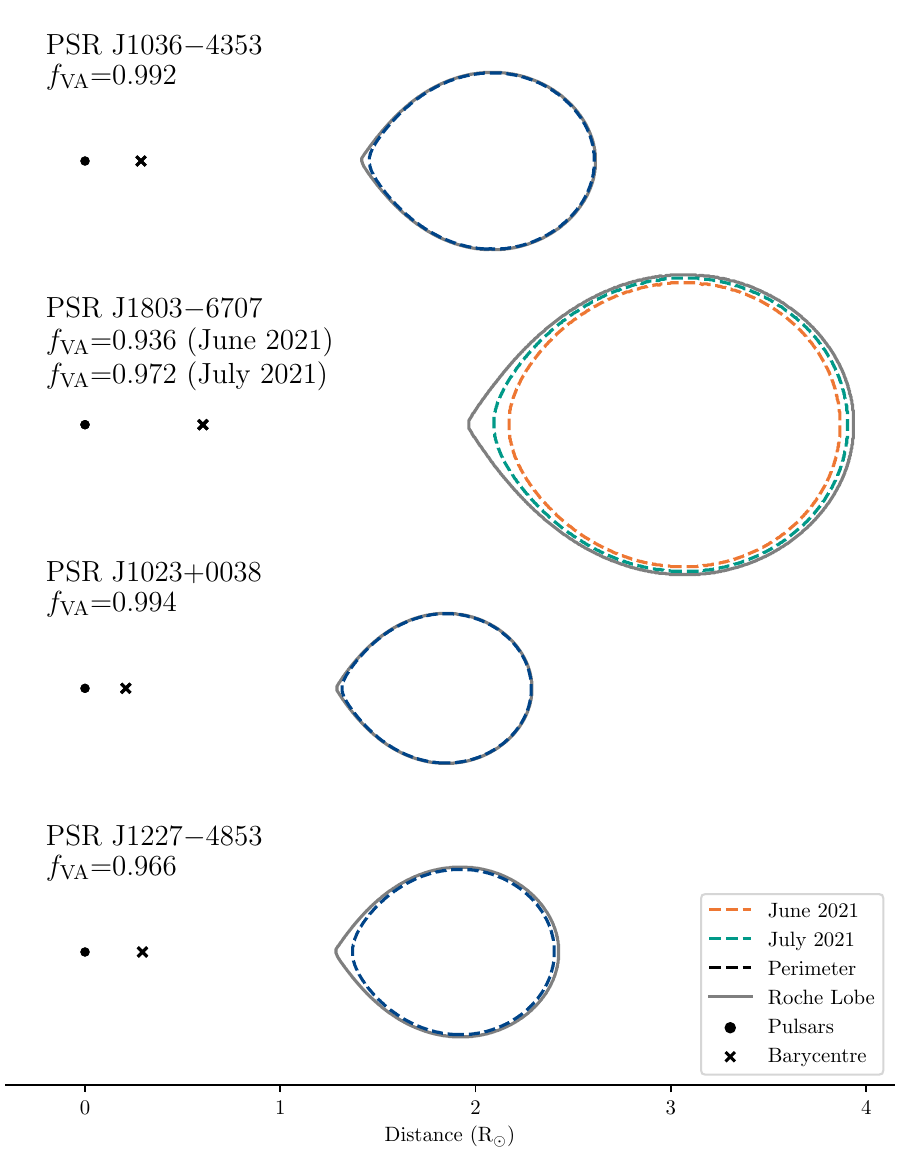}
    \caption{The projected outlines of PSRs~J1036$-$4353, J1803$-$6707, J1023$+$0038 and J1227$-$4853. The black filled dots are the position of the MSPs and the crosses mark the centre of mass of the binaries. The solid black lines show the Roche lobe around each companion. The dashed lines are the companion perimeters inferred from the best-fitting models. The green and orange dashed lines represent the companion of PSR~J1803$-$6707 in June and July 2021 respectively. \citet{Stringer2021} obtained the filling factors of the tMSPs, PSRs~J1023$+$0038 and J1227$-$4853, in the rotation-powered state, using an \texttt{Icarus} configuration that is similar to the pre-IGD prescription used in this work.}
    \label{fig:tmsp-fill-factor-plot}
\end{figure}

\subsection{PSR~J1803$-$6707 Companion Inflation}

A clear change in the light curve modulations of PSR~J1803$-$6707 is observed between June and July 2021, as shown in Figure~\ref{fig:lc_1803}. Within this time span, most orbital parameters, such as inclination, companion rotational period, and orbital separation, are expected to remain constant since the pulsar timing ephemeris does not change significantly and there is no evidence for mass transfer. However, the orbital periods of redback systems do vary significantly over time, believed to be due to changes in the internal structure of the companion star \citep{Voisin2020_quadrupole_moment}.  For PSR~J1803$-$6707, these variations in the orbital period have been monitored through the joint radio and $\gamma$-ray timing presented in \citet{Burgay2024+timing}. Between June 2021 and July 2021, the orbital period increased by around $\Delta p_{\rm orb} / p_{\rm orb} \approx 10^{-7}$, which would be consistent with a slight increase in the companion star's filling factor, although much larger variations (up to $|\Delta p_{\rm orb}|/p_{\rm orb} \approx 2\times10^{-6}$) are also seen during the \textit{Fermi} data for this pulsar. The relationship between the orbital period and the companion radius depends on the unknown ``apsidal motion'' constant \citep{Voisin2020_quadrupole_moment}, and so establishing a firm connection between these phenomena would require far more extensive optical monitoring. 

This variation in orbital period is very small and negligible for optical light curve modelling. Hence, the orbital parameters are kept fixed between the two epochs while modelling for the change in the light curve amplitude. This narrows down changing parameters to filling factor ($f_{\rm RL}$), base temperature ($T_{\rm base}$), irradiation temperature ($T_{\rm irr}$) and asymmetry-related parameters. Different combinations of these parameters are fitted to each data set while the other parameters are kept the same in all three datasets. The fitting with separate filling factor, irradiation temperature and the convection coefficient as separate parameter provides the highest Bayesian evidence. The constrained parameters are as follows: $K_{\rm c}$ = 244$^{+9}_{-10}$ km s$^{-1}$, inclination is approximately $53\pm2^\circ$, and the masses of the pulsar and companion are 1.7$\pm0.2$ and 0.44$^{+0.05}_{-0.04}$ M$_{\odot}$, respectively. We also evaluated the irradiation efficiency ($\varepsilon$), defined as the ratio of the irradiation energy rate to the spin-down luminosity, and found it remains constant within $1\sigma$. By contrast, the volume-averaged filling factor ($f_{\rm VA}$) increased by about 4 per~cent. This suggests that the change in incident irradiation may not be responsible for the companion's inflation. Instead, it is more likely that the evolution of the internal structure of the companion is the main cause underlying this inflation.

As shown in Figure~\ref{fig:tmsp-fill-factor-plot}, the volume-averaged filling factor of PSR~J1803$-$6707 varies from 0.936 and 0.972 between June and July 2021. By comparison, the volume-averaged filling factor of the two tMSPs PSR~J1023$+$0038 and PSR~J1227$-$4853 in their rotation-powered state were inferred to be 0.994 and 0.966, respectively, according to the best-fit model of their optical light curves \citep{Stringer2021}. It is unclear what their Roche-lobe filling factor is in the accretion state, but given that mass transfer is taking place it is presumably near unity. As such, one might speculate that if PSR~J1803$-$6707's companion had had a higher filling factor before its radius change -- at a similar level to that of the above two tMSPs during their rotation-powered state --, it could have experienced a tMSP state transition. The Roche-lobe would nearly filled and ensuing mass transfer could have formed an accretion disc, potentially leading to the disappearance of radio pulsations. Although this scenario has not been observed in PSR~J1803$-$6707, long-term monitoring could reveal further changes in which the system would undergo a successful state transition.

\section{Conclusion}
\label{sec:conclusion}

We searched for optical counterparts to eight new MSPs with MeerKAT. We obtained upper limits in optical bands for six binaries. Five of them are wide-orbit MSPs with white dwarf binaries which were expected to lie beyond detection limit. The sixth one, PSR~J1526$-$2744, is in a compact orbit with a low-mass companion which was expected to be a redback. The non-detection of a counterpart, however, most likely rules out this scenario in favour of being a rare example of an MSP having an extremely low-mass (ELM) white dwarf companion. The remaining two MSPs were expected to be redback systems, the nature of which was confirmed from detailed optical light curves. The modelling of these light curves and the spectroscopically-derived radial velocities we obtained allowed us to estimate the component masses. We found pulsar masses of 1.4$\pm0.1$ and 1.7$\pm0.2$ M$_{\odot}$ for PSRs J1036$-$4353 and J1803$-$6707, respectively.

In the optical follow-up campaign, preliminary radio localisation using \texttt{SeeKAT} proved to be useful in searching for the optical counterpart to the pulsar companions. This helped narrow down the search area to a few arcseconds only, which means that at most one or two stars would fall within the area of interest. The positional accuracy was subsequently confirmed by more accurate localisation from the following timing campaign.

We found that the post-irradiation gravity darkening prescription significantly improved the goodness-of-fit of our modelling in the case of PSR~J1036$-$4353. A phase shift in the modelled optical light curves compared to the observations using the pre-irradiation gravity darkening prescription provided the decisive evidence against this alternative. Thus, we conclude that high-energy irradiation from MSPs penetrates deeper than the layer of photosphere, in agreement with the proposal from \citet{Zilles2020} and \citet{Ginzburg2020}. More $\gamma$-ray photons from \textit{Fermi}-LAT should allow us to detect $\gamma$-ray eclipses and further constrain the inclination of PSR~J1036$-$4353.

The inferred change in the filling factor of PSR~J1803$-$6707 provides us with a significant clue regarding the possible triggering mechanism behind tMSP state transitions and the evolution of redbacks. Comparing its volume-averaged filling factor to that of known tMSPs suggests that this redback could have experienced a failed tMSP state-change episode due to the companion's envelope being insufficiently large enough to initiate Roche-lobe overflow. It is possible that further evolution could happen in the future, which would eventually lead to mass transfer. This source, along with other redbacks found to display light curves changing on timescales of weeks/months/years, requires long-term monitoring to better understand the mechanisms driving their evolution and possibly identify new tMSPs.

\section*{Acknowledgement}
R.P.B. acknowledges support from the European Research Council (ERC) under the European Union's Horizon 2020 research and innovation program (grant agreement No. 715051; Spiders).
J.S. acknowledges support from NASA grants 80NSSC23K1350 and 80NSSC22K1583, NSF grant AST-2205550, and the Packard Foundation.

This work was supported in part by the “Italian Ministry of Foreign Affairs and International Cooperation”, grant number ZA23GR03, under the project "RADIOMAP-Science and technology pathways to MeerKAT+: the Italian and South African synergy"

L.C., I.M. and K.V.S. are grateful for support from NSF grants AST-2107070 and AST-2205628, and NASA grants 80NSSC23K0497 and 80NSSC23K1247.

This work has made use of data from the European Space Agency (ESA) mission {\it Gaia} (\url{https://www.cosmos.esa.int/gaia}), processed by the {\it Gaia} Data Processing and Analysis Consortium (DPAC, \url{https://www.cosmos.esa.int/web/gaia/dpac/consortium}). Funding for the DPAC has been provided by national institutions, in particular the institutions participating in the {\it Gaia} Multilateral Agreement.

This research made use of \texttt{photutils}, an Astropy package for detection and photometry of astronomical sources \citep{Bradley2022}.

Based on observations collected at the European Southern Observatory under ESO programmes 0105.0695(B) and 0105.0700(A). We are grateful to the astronomy and technical staff support at La Silla Observatory who enabled the remote observation program and observers M.A. Kennedy and D. Mata Sanchez.

Based on observations obtained at the Southern Astrophysical Research (SOAR) telescope, which is a joint project of the Minist\'{e}rio da Ci\^{e}ncia, Tecnologia e Inova\c{c}\~{o}es (MCTI/LNA) do Brasil, the US National Science Foundation's NOIRLab, the University of North Carolina at Chapel Hill (UNC), and Michigan State University (MSU).

\section*{Data Availability}
The data from TRAPUM and ULTRACAM observations will be shared on reasonable request to the corresponding author. SOAR/Goodman observations are available from the NOIRLab Astro Data Archive \url{https://astroarchive.noirlab.edu/}.



\bibliographystyle{mnras_vanHack} 
\bibliography{trapum_paper} 




\appendix

\section{Posterior Distributions}

\begin{figure*}
    \centering
    \includegraphics[width=\textwidth]{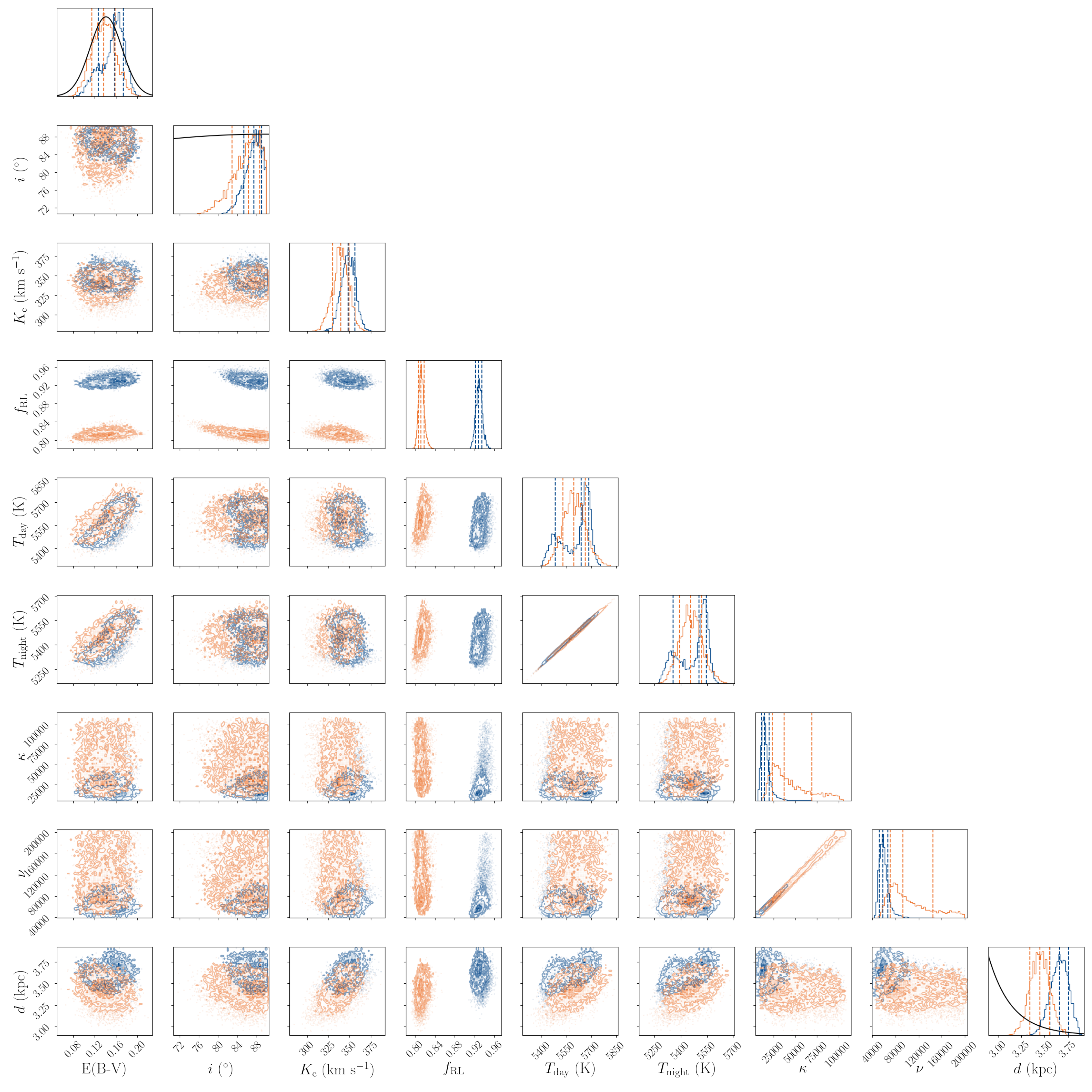}
    \caption{Posterior distribution of PSR~J1036$-$4353 using diffusion and convection models. Blue contours represent the Pre-IGD prescription, while orange contours represent the Post-IGD prescription.}
    \label{fig:corner_1036}
\end{figure*}

\begin{figure*}
    \centering
    \includegraphics[width=\textwidth]{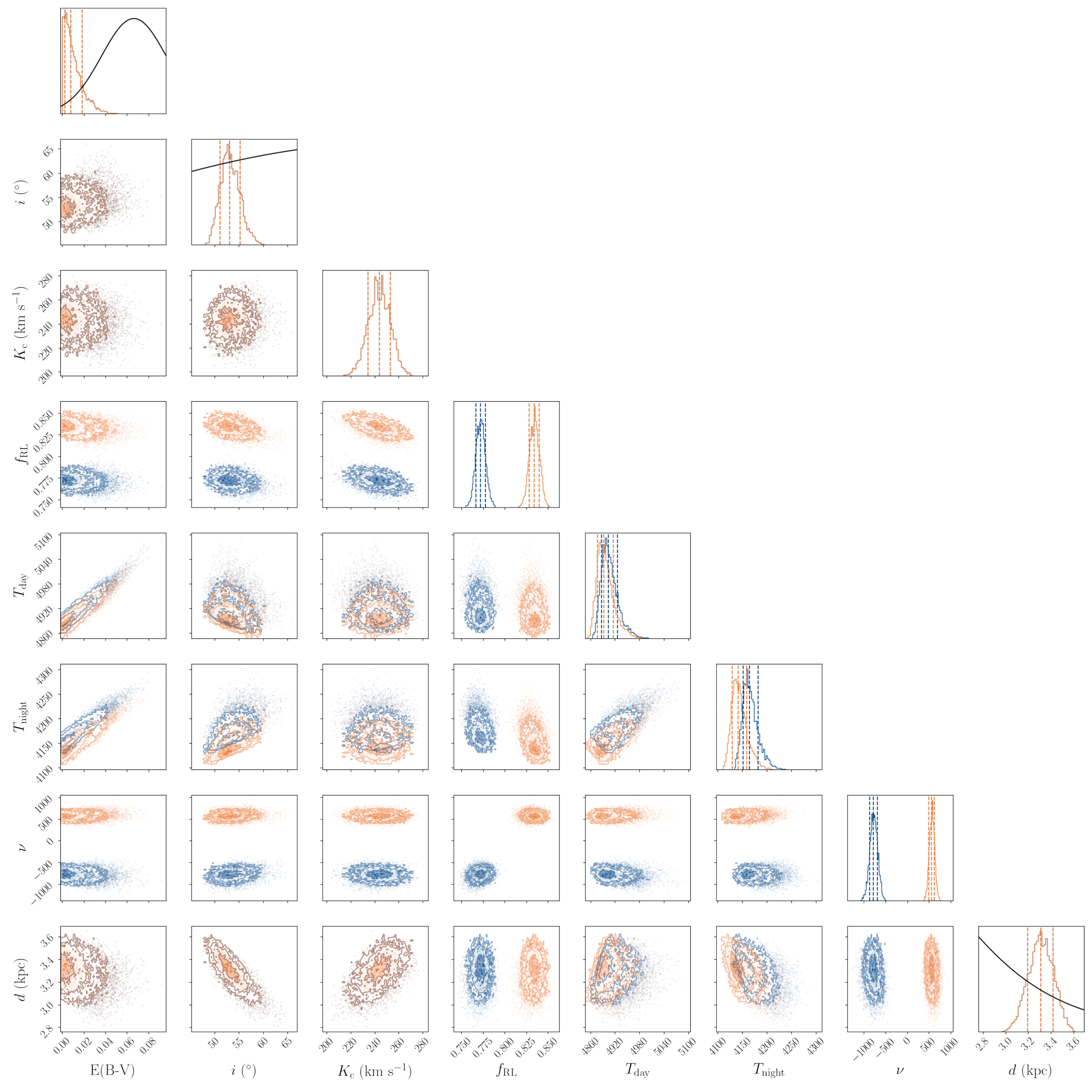}
    \caption{Posterior distribution of PSR~J1803$-$6707 using the best-fitting diffusion-and-convection model. Orange contours show fitted to the July 2021 data set, while blue contours represent the parameters fitted to the June 2021 data set. The E(B$-$V), inclination ($i$), radial velocity (K$_{\rm c}$) and distance (d) are joint-fitted parameters between the two data sets. Thus their contours are exactly overlapped.
}
    \label{fig:corner_1803}
\end{figure*}

\section{Parameters Tables}

\begin{table*}
\centering
\renewcommand{\arraystretch}{1.5}
\caption{Fitted and derived parameters from the models fitted to PSR~J1036$-$4353. The models use the gravity darkening exponent of 0.08.}
\label{tab:j1036_parameter_table}
\begin{tabular}{l|c|c|c|c|c|c|c|c}
\hline
Model & \multicolumn{2}{c}{Direct Heating} & \multicolumn{2}{c}{Convection} & \multicolumn{2}{c}{Diffusion+Convection} & \multicolumn{2}{c}{Spot} \\
\hline
Conv. Profile & & & \multicolumn{2}{c}{Uniform} & \multicolumn{2}{c}{Uniform} &  \\
\hline
 & Pre-IGD & Post-IGD & Pre-IGD & Post-IGD & Pre-IGD & Post-IGD & Pre-IGD & Post-IGD \\
\hline
\multicolumn{9}{c}{Fitted Parameters} \\
E(B$-$V) & $0.14_{-0.02}^{+0.02}$ & $0.14_{-0.02}^{+0.02}$ & $0.17_{-0.01}^{+0.01}$ & $0.14_{-0.02}^{+0.02}$ & $0.16_{-0.03}^{+0.02}$ & $0.14_{-0.02}^{+0.02}$ & $0.14_{-0.02}^{+0.02}$ & $0.14_{-0.02}^{+0.02}$ \\
$K_{\rm c}$ (km s$^{-1}$) & $341_{-8}^{+10}$ & $343_{-9}^{+9}$ & $340_{-10}^{+8}$ & $346_{-9}^{+10}$ & $348_{-9}^{+8}$ & $340_{-10}^{+9}$ & $336_{-10}^{+8}$ & $338_{-9}^{+9}$ \\
$d$ (kpc) & $3.42_{-0.1}^{+0.09}$ & $3.4_{-0.1}^{+0.1}$ & $3.64_{-0.09}^{+0.07}$ & $3.6_{-0.2}^{+0.2}$ & $3.66_{-0.1}^{+0.10}$ & $3.4_{-0.1}^{+0.1}$ & $3.40_{-0.09}^{+0.09}$ & $3.42_{-0.09}^{+0.09}$ \\
$i$ ($^\circ$) & $85_{-4}^{+3}$ & $85_{-4}^{+3}$ & $88_{-2}^{+1}$ & $77_{-6}^{+8}$ & $87_{-2}^{+2}$ & $86_{-4}^{+2}$ & $84_{-4}^{+3}$ & $86_{-4}^{+3}$ \\
$T_{\rm base}$ (K) & $5530_{-60}^{+50}$ & $5540_{-60}^{+60}$ & $5650_{-30}^{+20}$ & $5520_{-50}^{+40}$ & $5540_{-200}^{+60}$ & $5380_{-100}^{+90}$ & $5500_{-50}^{+50}$ & $5510_{-60}^{+60}$ \\
$T_{\rm irr}$ (K) & $3200_{-50}^{+50}$ & $3350_{-60}^{+60}$ & $4080_{-30}^{+20}$ & $3920_{-60}^{+60}$ & $4600_{-200}^{+200}$ & $5100_{-300}^{+600}$ & $2600_{-100}^{+200}$ & $2600_{-200}^{+200}$ \\
$f_{\rm RL}$ & $0.799_{-0.006}^{+0.008}$ & $0.796_{-0.006}^{+0.008}$ & $0.909_{-0.004}^{+0.005}$ & $0.82_{-0.02}^{+0.02}$ & $0.929_{-0.007}^{+0.006}$ & $0.812_{-0.005}^{+0.006}$ & $0.801_{-0.004}^{+0.007}$ & $0.798_{-0.005}^{+0.006}$ \\
$\nu$ (W K$^{-1}$ m$^{-2}$) & - & - & $35400_{-800}^{+700}$ & $31000_{-1000}^{+1000}$ & $58000_{-6000}^{+8000}$ & $90000_{-20000}^{+50000}$ & - & - \\
$\kappa$ (W K$^{-1}$ m$^{-2}$) & - & - & - & - & $14000_{-4000}^{+5000}$ & $40000_{-10000}^{+30000}$ & - & - \\
$\phi_{\rm spot}$ ($^\circ$) & - & - & - & - & - & - & $-52_{-4}^{+3}$ & $-50_{-5}^{+3}$ \\
$\theta_{\rm spot}$ ($^\circ$) & - & - & - & - & - & - & $142_{-4}^{+5}$ & $50_{-10}^{+90}$ \\
$R_{\rm spot}$ ($^\circ$) & - & - & - & - & - & - & $42_{-3}^{+3}$ & $45_{-4}^{+3}$ \\
$T_{\rm spot}$ (K) & - & - & - & - & - & - & $430_{-50}^{+80}$ & $360_{-60}^{+80}$ \\

\multicolumn{9}{c}{Band Offsets} \\
$r_s$ (mag) & $0.01_{-0.02}^{+0.02}$ & $0.01_{-0.02}^{+0.02}$ & $0.0_{-0.01}^{+0.01}$ & $0.01_{-0.02}^{+0.02}$ & $0.01_{-0.02}^{+0.02}$ & $0.01_{-0.02}^{+0.02}$ & $0.01_{-0.02}^{+0.02}$ & $0.01_{-0.02}^{+0.02}$ \\
$g_s$ (mag) & $-0.01_{-0.02}^{+0.02}$ & $-0.01_{-0.02}^{+0.02}$ & $-0.01_{-0.01}^{+0.01}$ & $-0.01_{-0.02}^{+0.02}$ & $-0.01_{-0.02}^{+0.01}$ & $-0.01_{-0.02}^{+0.02}$ & $-0.01_{-0.02}^{+0.01}$ & $-0.01_{-0.02}^{+0.02}$ \\
$u_s$ (mag) & $-0.01_{-0.04}^{+0.04}$ & $0.00_{-0.04}^{+0.04}$ & $0.06_{-0.02}^{+0.02}$ & $0.00_{-0.04}^{+0.03}$ & $0.04_{-0.07}^{+0.04}$ & $0.00_{-0.04}^{+0.04}$ & $0.01_{-0.04}^{+0.03}$ & $0.01_{-0.04}^{+0.03}$ \\
\multicolumn{9}{c}{Derived Parameters} \\
$q$ & $6.1_{-0.1}^{+0.2}$ & $6.1_{-0.2}^{+0.2}$ & $6.1_{-0.2}^{+0.1}$ & $6.2_{-0.2}^{+0.2}$ & $6.2_{-0.2}^{+0.1}$ & $6.1_{-0.2}^{+0.2}$ & $6.0_{-0.2}^{+0.1}$ & $6.1_{-0.2}^{+0.2}$ \\
$M_{\rm p}$ (M$_\odot$) & $1.47_{-0.10}^{+0.1}$ & $1.5_{-0.1}^{+0.1}$ & $1.44_{-0.1}^{+0.09}$ & $1.6_{-0.2}^{+0.2}$ & $1.5_{-0.1}^{+0.1}$ & $1.4_{-0.1}^{+0.1}$ & $1.4_{-0.1}^{+0.1}$ & $1.4_{-0.1}^{+0.1}$ \\
$M_{\rm c}$ (M$_\odot$) & $0.24_{-0.01}^{+0.01}$ & $0.24_{-0.01}^{+0.01}$ & $0.237_{-0.01}^{+0.009}$ & $0.26_{-0.02}^{+0.02}$ & $0.25_{-0.01}^{+0.01}$ & $0.24_{-0.01}^{+0.01}$ & $0.24_{-0.01}^{+0.01}$ & $0.24_{-0.01}^{+0.01}$ \\
$T_{\rm q1}$ (K) & $5470_{-60}^{+50}$ & $5480_{-60}^{+60}$ & $5660_{-30}^{+20}$ & $5540_{-50}^{+40}$ & $5620_{-200}^{+50}$ & $5570_{-70}^{+70}$ & $5570_{-60}^{+50}$ & $5570_{-60}^{+60}$ \\
$T_{\rm day}$ (K) & $5490_{-60}^{+60}$ & $5500_{-60}^{+60}$ & $5670_{-30}^{+20}$ & $5500_{-50}^{+40}$ & $5640_{-200}^{+50}$ & $5590_{-70}^{+70}$ & $5520_{-60}^{+50}$ & $5530_{-60}^{+60}$ \\
$T_{\rm q2}$ (K) & $5470_{-60}^{+50}$ & $5480_{-60}^{+60}$ & $5550_{-30}^{+20}$ & $5430_{-50}^{+40}$ & $5510_{-100}^{+40}$ & $5460_{-60}^{+60}$ & $5450_{-60}^{+50}$ & $5460_{-60}^{+50}$ \\
$T_{\rm night}$ (K) & $5390_{-60}^{+50}$ & $5390_{-60}^{+60}$ & $5530_{-30}^{+20}$ & $5390_{-50}^{+40}$ & $5500_{-100}^{+40}$ & $5460_{-60}^{+60}$ & $5400_{-60}^{+50}$ & $5410_{-60}^{+50}$ \\
$R_{\rm c}$ (R$_\odot$) & $0.455_{-0.007}^{+0.008}$ & $0.455_{-0.008}^{+0.008}$ & $0.476_{-0.008}^{+0.006}$ & $0.47_{-0.02}^{+0.02}$ & $0.485_{-0.007}^{+0.007}$ & $0.456_{-0.008}^{+0.008}$ & $0.452_{-0.007}^{+0.008}$ & $0.452_{-0.007}^{+0.007}$ \\
$v \sin{(i)}$ (km s$^{-1}$) & $88_{-1}^{+1}$ & $88_{-1}^{+1}$ & $93_{-2}^{+1}$ & $90_{-2}^{+2}$ & $94_{-1}^{+1}$ & $89_{-1}^{+1}$ & $87_{-1}^{+1}$ & $88_{-1}^{+1}$ \\
$f_{\rm VA}$ & $0.937_{-0.004}^{+0.005}$ & $0.935_{-0.004}^{+0.005}$ & $0.987_{-0.001}^{+0.001}$ & $0.951_{-0.009}^{+0.01}$ & $0.992_{-0.002}^{+0.001}$ & $0.945_{-0.003}^{+0.003}$ & $0.938_{-0.003}^{+0.004}$ & $0.936_{-0.003}^{+0.003}$ \\
$\varepsilon$ & $0.037_{-0.003}^{+0.003}$ & $0.045_{-0.004}^{+0.004}$ & $0.098_{-0.007}^{+0.005}$ & $0.090_{-0.010}^{+0.01}$ & $0.17_{-0.02}^{+0.03}$ & $0.24_{-0.06}^{+0.1}$ & $0.015_{-0.003}^{+0.004}$ & $0.016_{-0.004}^{+0.006}$ \\
\multicolumn{9}{c}{Goodness of fit} \\
$\ln{\rm (Z)}$           & -3300.9 & -3300.2 & -2364.1 & -2300.6 & -2341.8 & -2267.6 & -2277.4 & -2275.1 \\
N$_{\rm dof}$   & 3720 & 3720 & 3719 & 3719 & 3718 & 3718 & 3716 & 3716 \\
$\chi^{2}_{\rm \nu}$ & 2.146 & 2.146 & 1.304 & 1.268 & 1.290 & 1.247 & 1.248 & 1.248 \\
\hline
\end{tabular}
\end{table*}

\begin{table*}
\centering
\renewcommand{\arraystretch}{1.6}
\caption{Fitted and derived parameters from the models fitted to PSR~J1803$-$6707. The models shown in this table use only the Post-IGD prescription.}
\label{tab:parameters}
\begin{tabular}{l|c|c|c|c|c|c|c|c}
\hline
Model & \multicolumn{4}{c}{Convection} & \multicolumn{4}{c}{Diffusion+Convection} \\
\hline
Dataset Parameters & \multicolumn{2}{c}{$f_{\rm RL}$, $T_{\rm irr}$, and $\nu$} & \multicolumn{2}{c}{$f_{\rm RL}$, $T_{\rm base}$, $T_{\rm irr}$, and $\nu$} & \multicolumn{2}{c}{$f_{\rm RL}$, $T_{\rm irr}$, $\kappa$ and $\nu$} & \multicolumn{2}{c}{$f_{\rm RL}$, $T_{\rm base}$, $T_{\rm irr}$, $\kappa$ and $\nu$}\\
\hline
Dataset & June 2021 & July 2021 & June 2021 & July 2021 & June 2021 & July 2021 & June 2021 & July 2021 \\
\hline
\multicolumn{9}{c}{Fitted Parameters} \\
E(B$-$V) & \multicolumn{2}{c}{$0.013_{-0.009}^{+0.01}$} & \multicolumn{2}{c}{$0.014_{-0.010}^{+0.01}$} & \multicolumn{2}{c}{$0.008_{-0.006}^{+0.01}$} & \multicolumn{2}{c}{$0.012_{-0.005}^{+0.006}$} \\
$K_{\rm c}$ (km s$^{-1}$) & \multicolumn{2}{c}{$250_{-10}^{+10}$} & \multicolumn{2}{c}{$255_{-10}^{+10}$} & \multicolumn{2}{c}{$244_{-10}^{+9}$} & \multicolumn{2}{c}{$238_{-8}^{+6}$} \\
$d$ (kpc) & \multicolumn{2}{c}{$3.5_{-0.1}^{+0.1}$} & \multicolumn{2}{c}{$3.5_{-0.1}^{+0.1}$} & \multicolumn{2}{c}{$3.3_{-0.1}^{+0.1}$} & \multicolumn{2}{c}{$3.31_{-0.06}^{+0.06}$} \\
$i$ ($^\circ$) & \multicolumn{2}{c}{$52_{-2}^{+2}$} & \multicolumn{2}{c}{$51_{-2}^{+2}$} & \multicolumn{2}{c}{$53_{-2}^{+2}$} & \multicolumn{2}{c}{$52_{-1}^{+1}$} \\
$T_{\rm base}$ (K) & \multicolumn{2}{c}{$4270_{-20}^{+20}$} & $4270_{-20}^{+20}$ & $4270_{-20}^{+20}$ & \multicolumn{2}{c}{$4250_{-10}^{+20}$} & $4240_{-10}^{+10}$ & $4256_{-9}^{+10}$ \\
$T_{\rm irr}$ (K) & $4650_{-40}^{+40}$ & $4720_{-40}^{+40}$ & $4670_{-40}^{+50}$ & $4740_{-40}^{+50}$ & $4740_{-40}^{+40}$ & $4670_{-30}^{+40}$ & $4820_{-30}^{+30}$ & $4690_{-20}^{+20}$ \\
$f_{\rm RL}$ & $0.822_{-0.005}^{+0.005}$ & $0.843_{-0.005}^{+0.005}$ & $0.817_{-0.006}^{+0.006}$ & $0.846_{-0.006}^{+0.005}$ & $0.772_{-0.006}^{+0.006}$ & $0.834_{-0.006}^{+0.006}$ & $0.769_{-0.005}^{+0.004}$ & $0.830_{-0.005}^{+0.005}$ \\
$\nu$ (W K$^{-1}$ m$^{-2}$) & $-580_{-80}^{+80}$ & $570_{-60}^{+70}$ & $-610_{-80}^{+80}$ & $570_{-70}^{+70}$ & $-770_{-90}^{+90}$ & $570_{-60}^{+60}$ & $-800_{-80}^{+80}$ & $570_{-50}^{+50}$ \\
$\kappa$ (W K$^{-1}$ m$^{-2}$) & \multicolumn{2}{c}{-} & \multicolumn{2}{c}{-} & $1500_{-100}^{+100}$ & $20_{-20}^{+30}$ & $2100_{-200}^{+200}$ & $20_{-20}^{+30}$ \\
\multicolumn{9}{c}{Band Offsets} \\
$i_s$ (mag) & $0.01_{-0.02}^{+0.02}$ & $-0.02_{-0.02}^{+0.02}$ & $0.01_{-0.02}^{+0.02}$ & $-0.02_{-0.02}^{+0.02}$ & $0.00_{-0.02}^{+0.02}$ & $-0.01_{-0.02}^{+0.02}$ & $0.00_{-0.02}^{+0.02}$ & $-0.01_{-0.02}^{+0.02}$ \\
$g_s$ (mag) & $0.03_{-0.01}^{+0.01}$ & $0.03_{-0.01}^{+0.02}$ & $0.03_{-0.02}^{+0.02}$ & $0.03_{-0.02}^{+0.02}$ & $0.02_{-0.02}^{+0.02}$ & $0.01_{-0.02}^{+0.02}$ & $0.02_{-0.02}^{+0.02}$ & $0.01_{-0.02}^{+0.02}$ \\
$u_s$ (mag) & $-0.04_{-0.01}^{+0.01}$ & $-0.02_{-0.01}^{+0.01}$ & $-0.04_{-0.01}^{+0.01}$ & $-0.02_{-0.02}^{+0.02}$ & $-0.03_{-0.01}^{+0.01}$ & $-0.002_{-0.009}^{+0.009}$ & $-0.03_{-0.01}^{+0.01}$ & $-0.004_{-0.009}^{+0.009}$ \\
\multicolumn{9}{c}{Derived Parameters} \\
$q$ & \multicolumn{2}{c}{$4.2_{-0.2}^{+0.2}$} & \multicolumn{2}{c}{$4.2_{-0.2}^{+0.2}$} & \multicolumn{2}{c}{$4.0_{-0.2}^{+0.2}$} & \multicolumn{2}{c}{$3.92_{-0.1}^{+0.10}$} \\
$M_{\rm p}$ (M$_\odot$) & \multicolumn{2}{c}{$2.0_{-0.2}^{+0.3}$} & \multicolumn{2}{c}{$2.1_{-0.3}^{+0.3}$} & \multicolumn{2}{c}{$1.7_{-0.2}^{+0.2}$} & \multicolumn{2}{c}{$1.7_{-0.1}^{+0.1}$} \\
$M_{\rm c}$ (M$_\odot$) & \multicolumn{2}{c}{$0.48_{-0.05}^{+0.05}$} & \multicolumn{2}{c}{$0.50_{-0.05}^{+0.05}$} & \multicolumn{2}{c}{$0.44_{-0.04}^{+0.05}$} & \multicolumn{2}{c}{$0.44_{-0.02}^{+0.02}$} \\
$T_{\rm q1}$ (K) & $4450_{-20}^{+20}$ & $4480_{-20}^{+30}$ & $4450_{-20}^{+30}$ & $4480_{-20}^{+30}$ & $4440_{-10}^{+20}$ & $4420_{-10}^{+20}$ & $4449_{-9}^{+10}$ & $4429_{-9}^{+10}$ \\
$T_{\rm day}$ (K) & $4900_{-20}^{+30}$ & $4920_{-20}^{+30}$ & $4910_{-20}^{+40}$ & $4930_{-20}^{+40}$ & $4900_{-20}^{+20}$ & $4890_{-20}^{+20}$ & $4910_{-10}^{+20}$ & $4910_{-10}^{+20}$ \\
$T_{\rm q2}$ (K) & $4470_{-20}^{+20}$ & $4460_{-20}^{+20}$ & $4480_{-20}^{+30}$ & $4460_{-20}^{+30}$ & $4460_{-10}^{+20}$ & $4400_{-10}^{+20}$ & $4472_{-9}^{+10}$ & $4409_{-9}^{+10}$ \\
$T_{\rm night}$ (K) & $4160_{-20}^{+20}$ & $4160_{-20}^{+20}$ & $4160_{-20}^{+20}$ & $4150_{-20}^{+20}$ & $4160_{-10}^{+20}$ & $4140_{-10}^{+20}$ & $4156_{-9}^{+10}$ & $4145_{-9}^{+10}$ \\
$R_{\rm c}$ (R$_\odot$) & $0.75_{-0.02}^{+0.02}$ & $0.76_{-0.02}^{+0.02}$ & $0.76_{-0.03}^{+0.02}$ & $0.77_{-0.03}^{+0.03}$ & $0.70_{-0.02}^{+0.02}$ & $0.73_{-0.02}^{+0.02}$ & $0.70_{-0.01}^{+0.01}$ & $0.73_{-0.01}^{+0.01}$ \\
$v \sin{(i)}$ (km s$^{-1}$) & $79_{-2}^{+2}$ & $80_{-2}^{+2}$ & $79_{-2}^{+1}$ & $80_{-2}^{+1}$ & $75_{-1}^{+1}$ & $78_{-1}^{+1}$ & $73.5_{-1}^{+1.0}$ & $76.5_{-1}^{+1.0}$ \\
$f_{\rm VA}$ & $0.949_{-0.003}^{+0.003}$ & $0.960_{-0.002}^{+0.003}$ & $0.947_{-0.003}^{+0.003}$ & $0.962_{-0.003}^{+0.003}$ & $0.918_{-0.004}^{+0.004}$ & $0.956_{-0.003}^{+0.003}$ & $0.916_{-0.003}^{+0.003}$ & $0.953_{-0.003}^{+0.002}$ \\
$\varepsilon$ & $0.21_{-0.02}^{+0.02}$ & $0.22_{-0.02}^{+0.02}$ & $0.22_{-0.02}^{+0.03}$ & $0.23_{-0.02}^{+0.03}$ & $0.21_{-0.02}^{+0.02}$ & $0.20_{-0.02}^{+0.02}$ & $0.22_{-0.01}^{+0.01}$ & $0.20_{-0.01}^{+0.01}$ \\
\multicolumn{9}{c}{Goodness of fit} \\
$\ln{\rm (Z)}$           & \multicolumn{2}{c}{-6656.2} & \multicolumn{2}{c}{-6661.3} & \multicolumn{2}{c}{-6622.2} & \multicolumn{2}{c}{-6629.6} \\
N$_{\rm dof}$   & \multicolumn{2}{c}{12487} & \multicolumn{2}{c}{12486} & \multicolumn{2}{c}{12485} & \multicolumn{2}{c}{12484} \\
$\chi^{2}_{\rm \nu}$ & \multicolumn{2}{c}{1.062} & \multicolumn{2}{c}{1.062} & \multicolumn{2}{c}{1.053} & \multicolumn{2}{c}{1.053} \\
\hline
\end{tabular}
\end{table*}


\bsp    
\label{lastpage}
\end{document}